\documentclass[useAMS]{mn2e}

\usepackage{graphicx}

\title[Rotation of Sirius A]
{On the rotational velocity of Sirius A
}

\author[Y. Takeda]
{Yoichi Takeda\thanks{E-mail:
ytakeda@js2.so-net.ne.jp}\footnotemark[0] 
\\
11-2 Enomachi, Naka-ku, Hiroshima-shi 730-0851, Japan\\
}
\begin{document}

\date{Accepted 2020 September 11. Received 2020 September 11; in original form 2020 August 10}


\maketitle

\label{firstpage}

\begin{abstract}
With an aim of getting information on the equatorial rotation velocity ($v_{\rm e}$) of 
Sirius~A separated from the inclination effect ($\sin i$), a detailed profile analysis 
based on the Fourier transform technique was carried out for a large number of spectral lines, 
while explicitly taking into account the line-by-line differences in the centre--limb 
behaviours and the gravity darkening effect (which depend on the physical properties of 
each line) based on model calculations. 
The simulations showed that how the 1st-zero frequencies ($q_{1}$) of Fourier transform 
amplitudes depends on $v_{\rm e}$ is essentially determined by the temperature-sensitivity 
parameter ($K$) differing from line to line, and that Fe~{\sc i} lines (especially those 
of very weak ones) are more sensitive to $v_{\rm e}$ than Fe~{\sc ii} lines.   
The following conclusions were drawn by comparing the theoretical and observed 
$q_{1}$ values for many Fe~{\sc i} and Fe~{\sc ii} lines: (1) The projected rotational
velocity ($v_{\rm e}\sin i$) for Sirius~A is fairly well established at 
$16.3 (\pm 0.1)$~km~s$^{-1}$  by requiring that both Fe~{\sc i} and Fe~{\sc ii} 
lines yield consistent results. (2) Although precise separation of 
$v_{\rm e}$ and $i$ is difficult, $v_{\rm e}$ is concluded to be
in the range of $16 \le v_{\rm e} \la$~30--40~km~s$^{-1}$, which corresponds to
$25^{\circ} \la i \le 90^{\circ}$. Accordingly, Sirius~A is an intrinsically 
slow rotator for an A-type star, being consistent with its surface chemical peculiarity.
\end{abstract}

\begin{keywords}
stars: atmospheres --- stars: chemically peculiar --- stars: early-type 
stars: individual (Sirius) --- stars: rotation
\end{keywords}

\section{Introduction}
 
Sirius~A (= $\alpha$~CMa = HD~48915 = HR~2491 = HIP~32349), the brightest star 
in the night sky, is a main-sequence star of A1V type, which constitutes a visual 
binary system of 50~yr period along with the DA white dwarf companion Sirius~B.
This star is rather unusual in the sense that it shows comparatively sharp lines 
(projected rotational velocity $v_{\rm e}\sin i$ being on the order of 
$\sim$~15--20~km~s$^{-1}$, where $v_{\rm e}$ is the equatorial rotation velocity 
and $i$ is the angle of rotational axis relative to the line of sight) despite 
that $v_{\rm e}\sin i$ values of most A-type dwarfs typically range $\sim$~100--300~km~s$^{-1}$
(see, e.g., Fig.~5 in Abt \& Morrell 1995). 

In this context, it is worth referring to the case of Vega, another popular A-type star. 
Although it is a similar sharp-line star ($v_{\rm e}\sin i \sim 20$~km~s$^{-1}$), 
this is nothing but a superficial effect caused by very small $\sin i$ (i.e., 
fortuitously seen nearly pole-on). That Vega is actually a pole-on rapid rotator 
with large $v_{\rm e}$ ($\sim 200$~km~s$^{-1}$) showing an appreciable gravity-darkening 
has been well established in various ways such as line profile analysis (e.g., 
Takeda, Kawanomoto \& Ohishi 2008), spectropolarimetric observation of magnetic 
fields (e.g., Petit et al. 2010) and interferometric observations 
(e.g., Monnier et al. 2012).

In contrast, Sirius~A seems unlikely to be such a pole-on-seen rapid rotator as argued 
by Petit et al. (2011), because it is classified as a chemically peculiar star
(metallic-lined A-type star or Am star) showing specific surface abundance anomalies
characterised by enrichments of Fe group or s-process elements as well as deficiencies 
of some species such as Ca or Sc (see, e.g., Landstreet 2011 and the references therein). 
The point is that these Am phenomena are seen only in comparatively low $v_{\rm e}\sin i$ 
stars ($v_{\rm e} < 120$~km~s$^{-1}$; cf. Abt 2009), by which the possibility of Sirius~A 
intrinsically rotating very rapidly like other A-stars may be ruled out.  

Yet, such a broad restriction on $v_{\rm e}$ (widely ranging from $\sim 10$ to
$\sim 100$~km~s$^{-1}$ is far from sufficient. More practical information with 
$v_{\rm e}$ (and $i$) of more limited range would be needed, in order for better 
understanding of the Sirius system (e.g., whether the axis of rotation is aligned 
with that of the orbital motion). Unfortunately, such a trial seems to have never 
been tried so far, despite that quite a few number of (unseparated) $v_{\rm e}\sin i$ 
determinations for this star are published as summarised in Table~1. 

\setcounter{table}{0}
\begin{table*}
\begin{minipage}{180mm}
\small
\caption{Previous determinations of the projected rotational velocity for Sirius~A.}
\begin{center}
\begin{tabular}{ccl}\hline
\hline
Authors  &   $v_{\rm e}\sin i$ & Determination method  \\
         &  (km~s$^{-1}$)   &     \\
\hline
Smith (1976)    & 17 & Fourier transform (fit of the main/side lobe, Fe~{\sc ii} 4923 line) \\
Deeming (1977)  & 16.9  & Fourier+Bessel transform (Fe~{\sc ii} 4923) \\
Milliard et al. (1977) & 11 & Fourier transform (Fe~{\sc ii} 1362 and Cl~{\sc i} 1363) \\
Dravins et al. (1990) &  15.3 & Fourier transform (3 Fe~{\sc i} and 1 Fe~{\sc ii} lines) \\
Apt \& Morrell (1995) & 15 & half widths measured by Gaussian fitting (Mg~{\sc ii} 4481 and Fe~{\sc i} 4476)  \\
Landstreet (1998) & 16.5 & synthetic profile fitting (several Fe~{\sc ii} and Cr~{\sc ii} lines in 4610--4640~\AA) \\
Royer et al. (2002) & 16 & Fourier transform (1st zeros for a number of lines in the blue region) \\
D\'{\i}az et al. (2011) & 16.7 & Fourier transform (1st zeros, application of cross correlation function)\\
Takeda et al. (2012) & 16.6 & synthetic spectrum fitting (in the range of 6146--6163 \AA) \\
\hline
\end{tabular}
\end{center}
\end{minipage}
\end{table*}

This situation motivated the author to challenge this task based on the 
previous experiences of separating $v_{\rm e}$ and $i$ for rapidly rotating stars, 
where the spectral line profiles (having different sensitivities to the 
gravitational darkening effect) were analysed with the help of model simulations: 
(i) study of Vega by Takeda et al. (2008) using a number of weak lines of neutral 
and ionised species, and (ii) investigation on 6 rapid rotators of late-B type 
by Takeda, Kawanomoto \& Ohishi (2017) based on the Fourier transform analysis of 
the He~{\sc i} 4471 + Mg~{\sc ii} 4481 feature. 

However, the problem should be more difficult compared to those previous cases
as Sirius~A is expected to rotate more slowly ($v_{\rm e} < 120$~km~s$^{-1}$), 
and it is not clear whether the gravity-darkening effect (and its delicate
difference depending on lines) could be sufficiently large to be detected. 
Accordingly, the following strategy was adopted in this study:
\begin{itemize}
\item
In light of inevitable fluctuations or ambiguities due to observational errors,
as many suitable spectral lines as possible (from very weak to strong saturated 
lines) are to be employed, so that statistically meaningful discussions
may be possible.
\item
Instead of direct fitting with the modelled profile in the usual wavelength domain
(such as done in Takeda et al. 2008), specific zero frequencies obtained by Fourier 
transforming of line profiles are used (as done by Takeda et al. 2017) for comparing 
observables with the modelling results, which are measurable with high precision
and advantageous for detecting very subtle profile differences. 
\item
In order to understand which kind of line is how sensitive to changing $v_{\rm e}$, 
the behaviours of Fourier transform parameters (zero frequencies etc.) in terms of
the line properties (e.g., line strength, species, excitation potential, 
temperature-sensitivity) are investigated.
\end{itemize}

\section{Observed line profiles and transforms}

\subsection{Observational data}

The spectroscopic observation of Sirius~A was carried out on 2008 October 7 ad 8 (UT)
by using HIDES (HIgh Dispersion Echelle Spectrograph) placed at the coud\'{e} focus 
of the 188 cm reflector at Okayama Astrophysical Observatory.
Equipped with three mosaicked 4K$\times$2K CCD detectors at the camera focus, 
echelle spectra covering 4100--7800~\AA\ (in the mode of red cross-disperser) 
with a resolving power of $R \sim 100000$ (corresponding to the slit width 
of 100~$\mu$m) were obtained. 

The reduction of the spectra (bias subtraction, flat-fielding, 
scattered-light subtraction, spectrum extraction, wavelength 
calibration, and continuum normalisation) was performed by using 
the ``echelle'' package of the software IRAF\footnote{
IRAF is distributed by the National Optical Astronomy Observatories,
which is operated by the Association of Universities for Research
in Astronomy, Inc. under cooperative agreement with the National 
Science Foundation.} in a standard manner. 

Very high S/N ratios (typically around $\sim 2000$ on the average) could be 
accomplished in the final spectra by co-adding 25 frames (each with exposure 
time of 15--60~s), as shown in Fig.~1 (S/N $\sim \sqrt{count}$).   

\setcounter{figure}{0}
\begin{figure*}
\begin{minipage}{100mm}
\includegraphics[width=10.0cm]{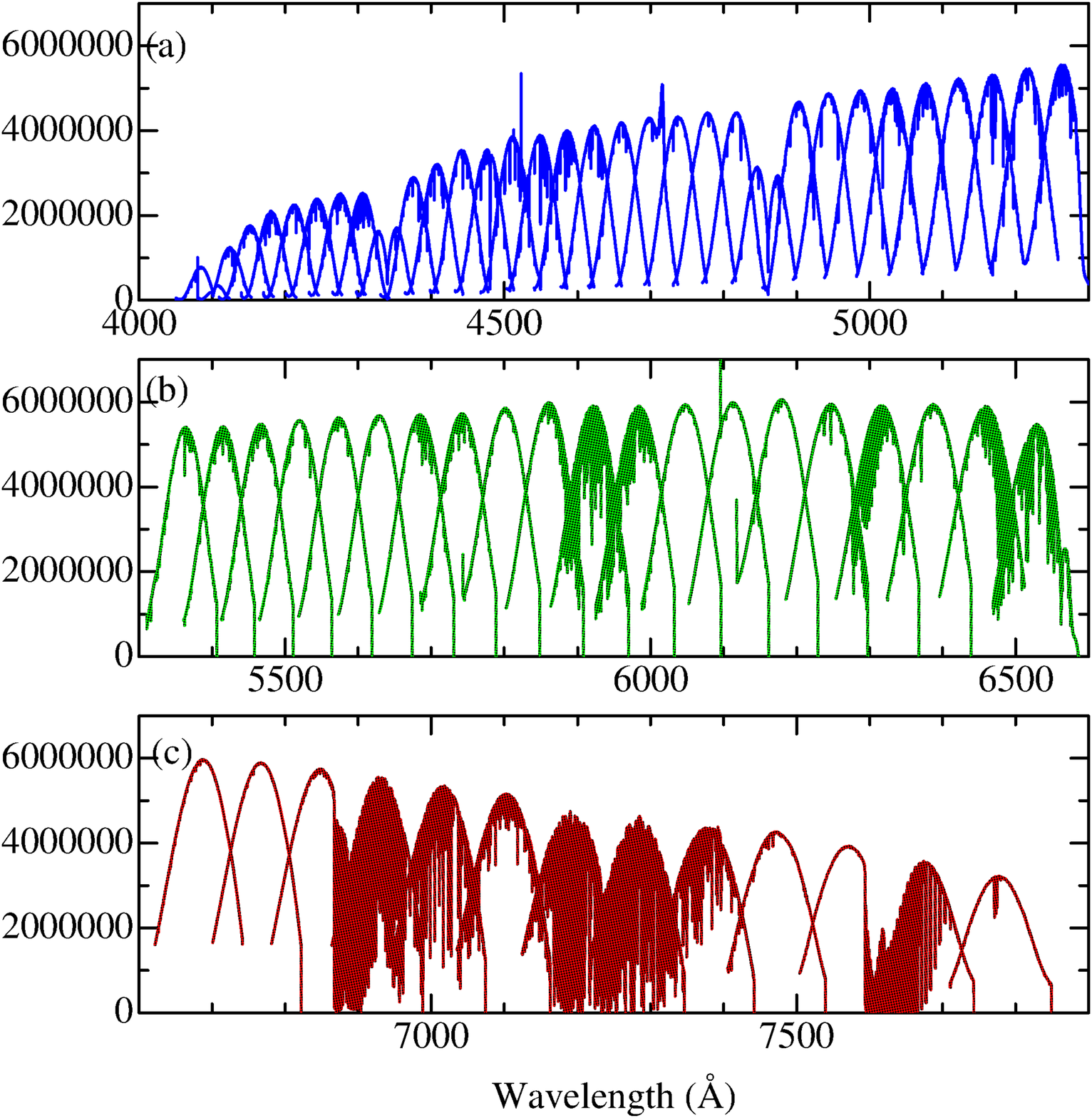}
\caption{
Panels (a), (b), and (c) show the distributions of accumulated photoelectron 
counts of CCD for the final spectra of Sirius~A, each corresponding to three 
wavelength regions (4100--5300/5300--6600/6600--7900~\AA) comprising 32/20/13 
orders, respectively. 
Note that the S/N ratio can be estimated as S/N $\sim \sqrt{count}$ in the present 
photon-noise-limited case.
The spectra in each of the echelle orders show characteristic distributions of 
the blaze function. 
}
\label{fig1}
\end{minipage}
\end{figure*}

\subsection{Line selection}

The candidate lines to be analysed were first sorted out by inspecting 
the observed spectral feature while comparing it with the calculated strengths 
of neighbourhood lines as well as the synthesised theoretical spectrum, where
the simulation was done by including the atomic lines taken from VALD database
(Ryabchikova et al. 2015). Here, the following selection criteria were adopted.
\begin{itemize}
\item
The line should not be severely contaminated by blending (due to other lines or telluric 
lines). Accordingly, it is checked that the observed wavelength at the flux minimum 
(i.e., line centre) reasonably coincide with the expected line wavelength.
\item
The feature should be dominated by only one line component
as judged by comparing the theoretical simulation results.
(For example, the Mg~{\sc ii} 4481 line was not adopted because it comprises two 
doublet components.) 
\item
The contamination of profile wings by neighbourhood lines can often be a problem. 
After the two limiting wavelengths ($\lambda_{1}$, $\lambda_{2}$; defining the usable
portion of the profile) and the continuum level\footnote{
Although the original spectrum was already normalised by using the IRAF task ``continuum'', 
a slight adjustment of the local continuum position was often necessary for this 
kind of profile analysis, which should be done interactively.} ($F_{\rm cont}$) 
were determined by eye-inspection, it was required that at least either one of the 
limits (say, $\lambda_{1}$) should reach the continuum level (i.e., 
$F(\lambda_{1}) \simeq F_{\rm cont}$), and that the departure should not be 
significant for the other limit (say, $\lambda_{2}$) even if $F(\lambda_{2})$ 
is somewhat below the continuum due to blending.    
\end{itemize}

As a result, 571 lines of 21 elements were selected (C~{\sc i}, N~{\sc i},
O~{\sc i}, Mg~{\sc i/ii}, Al~{\sc ii}, Si~{\sc i/ii}, S~{\sc i/ii}, Ca~{\sc i},
Sc~{\sc ii}, Ti~{\sc ii}, V~{\sc ii}, Cr~{\sc i/ii}, Mn~{\sc i/ii}, Fe~{\sc i/ii},
Co~{\sc i/ii}, Ni~{\sc i/ii}, Zn~{\sc i}, Sr~{\sc ii}, Y~{\sc ii}, Ba~{\sc ii}, 
and Ce~{\sc ii}). The observed profiles of these lines are graphically 
displayed in Fig.~2 and Fig.~3. The basic atomic data and the profile data of all
571 lines are also presented in ``measuredlines.dat'' and ``obsprofiles.dat'' 
of the online material, respectively.

\setcounter{figure}{1}
\begin{figure*}
\begin{minipage}{140mm}
\includegraphics[width=14.0cm]{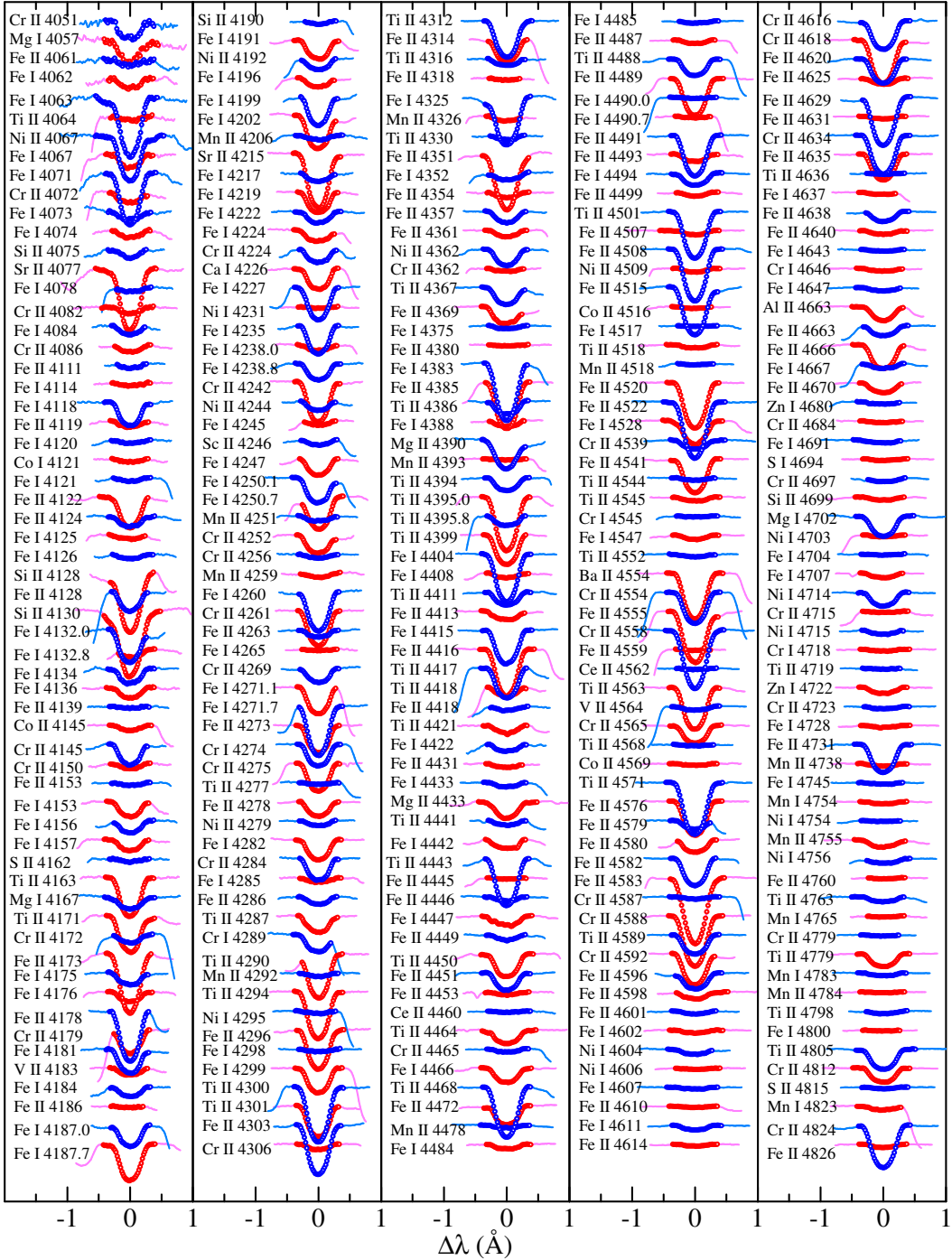}
\caption{
Observed spectra of finally selected lines (shown here are 300 lines out of the 
total 571 lines), which are arranged in the increasing order of wavelength. 
The original spectral features (normalised flux plotted against the wavelength 
displacement relative to the line centre) are shown by lines, while the wavelength 
portions [$\lambda_{1}$, $\lambda_{2}$] used for calculating the Fourier transforms 
are depicted by symbols. Each spectrum is shifted by 0.1 (10\% of the continuum level) 
relative to the adjacent one.
}
\label{fig2}
\end{minipage}
\end{figure*}

\setcounter{figure}{2}
\begin{figure*}
\begin{minipage}{140mm}
\includegraphics[width=14.0cm]{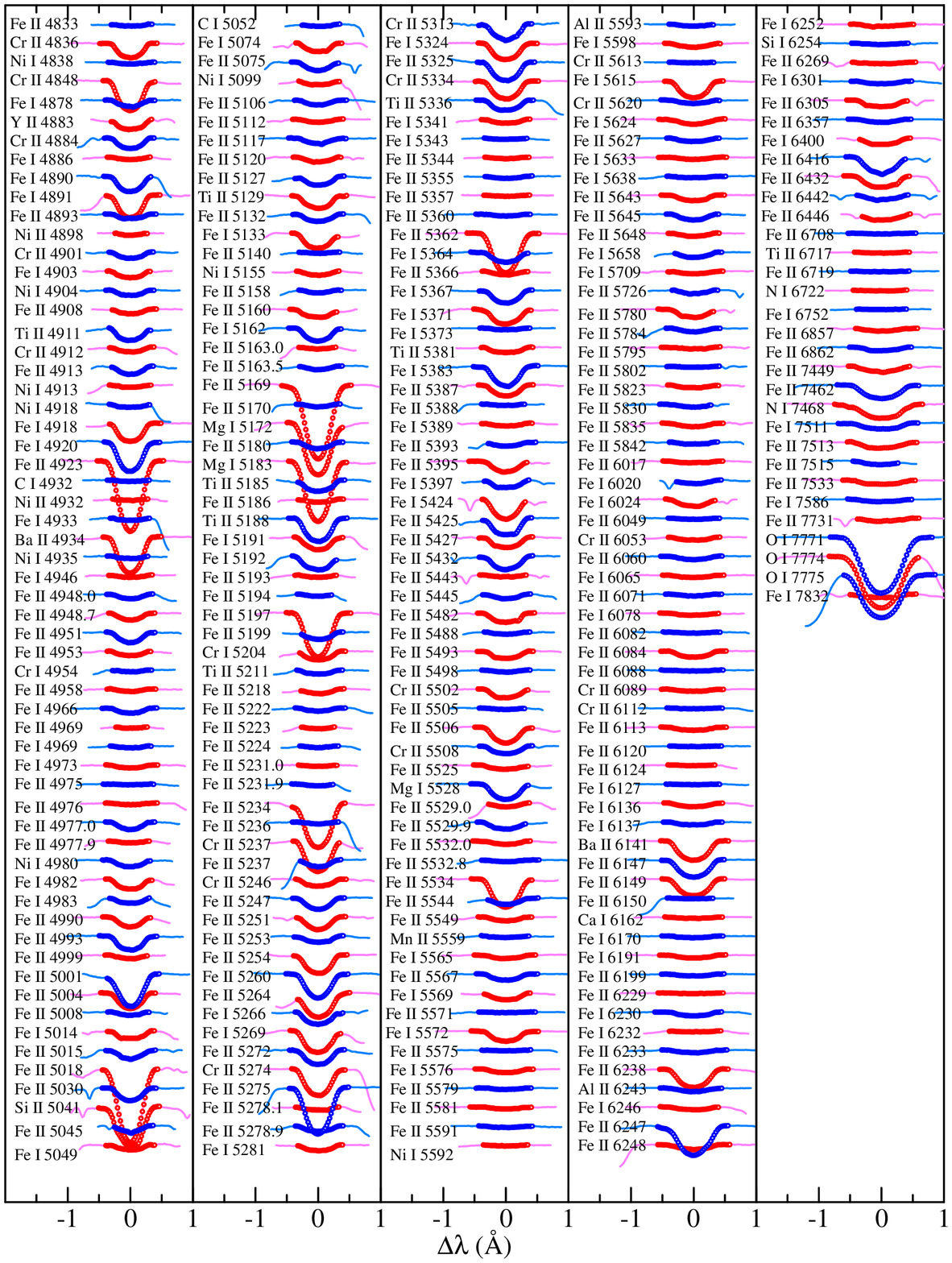}
\caption{
Observed spectra of finally selected lines (271 lines out of the total 571 lines). 
Otherwise, the same as Fig.~2.
}
\label{fig3}
\end{minipage}
\end{figure*}

The equivalent width ($W$) of each line was evaluated by the non-linear 
least-squares fitting procedure, which fits the observed profile ($F(\lambda)$) 
with a specifically devised function (constructed by convolving the rotational 
broadening function with the Gaussian function in an appropriate fashion) 
while making use of $\lambda_{1}$, $\lambda_{2}$, and $F_{\rm cont}$ as input parameters.
The resulting $W$ values range widely from $\sim 1$~m\AA\ (very weak line) to 
$\sim 200$~m\AA\ (strong saturated line).
Besides, the line-by-line abundances ($A$: logarithmic number abundance with 
the usual normalisation of $A_{\rm H} = 12$) were calculated from these $W$ values
by using the Kurucz's (1993) WIDTH9 program and ATLAS9 model atmosphere with $T_{\rm eff} = 10000$~K, 
$\log g = 4.30$, [M/H] = +0.5~dex, and $v_{\rm t} = 2$~km~s$^{-1}$ (which are the typical 
values of effective temperature, surface gravity, metallicity, and microturbulence for 
Sirius~A, respectively; cf. Landstreet 2011 and the references therein). 
Such obtained $W$ and $A$ results are also given in ``measuredlines.dat''.

\subsection{Fourier transform and zero frequencies}

The Fourier transform of the line depth profile $D(\lambda)$,  
\begin{equation}
D(\lambda) \equiv 1 - F(\lambda)/F_{\rm cont}
\end{equation} 
is expressed as
\begin{equation}
  d(\sigma) \equiv \int_{-\infty}^{\infty} D(\lambda)\exp(2\pi i\sigma\lambda) {\rm d}\lambda
\end{equation}
 (see, e.g., Gray 2005).
The zero frequencies of the Fourier transform amplitude ($|d(\sigma)|$) are measurable as
$\sigma_{1}$, $\sigma_{2}$, ..., which are called as 1st-zero, 2nd-zero, ..., respectively.
The related quantities which will be later referred to are the heights of two lobes: 
(i) the main lobe height ($h^{0}$) is $|d(\sigma)|$ at $\sigma = 0$ (equal to the equivalent width $W$) 
and (ii) the side lobe height ($h^{1}$) is the peak of $|d(\sigma)|$ between $\sigma_{1}$ and $\sigma_{2}$.  
Since the Fourier frequency $\sigma$ (in unit of \AA$^{-1}$) depends on the line wavelength
and not useful for comparing the cases of different lines, scaled frequency $q$  
is used in this study for zero frequencies as $q_{1}$, $q_{2}$, ... (in unit of km$^{-1}$s), 
which is defined as $q \equiv \sigma (\lambda/299792.5)$ (where $\lambda$ is in \AA).
 
As such, the Fourier transform $d(\sigma)$ of each line was calculated from its profile 
$D(\lambda)$ according to Eq.~2, where the numerical integration over the wavelength was 
done at [$\lambda_{1}$, $\lambda_{2}$].
The resulting values of $q_{1}$, $q_{2}$, $h^{0}$, $h^{1}$ measured from $|d(\sigma)|$ are
summarised in ``measuredlines.dat''.   
As typical examples, Fig.~4 illustrates the profiles ($F(\lambda)$) along with the corresponding 
transforms ($|d(\sigma)|$) for 10 representative Fe lines of various strengths (5 Fe~{\sc i} 
lines and 5 Fe~{\sc ii} lines). 

\setcounter{figure}{3}
\begin{figure*}
\begin{minipage}{160mm}
\includegraphics[width=16.0cm]{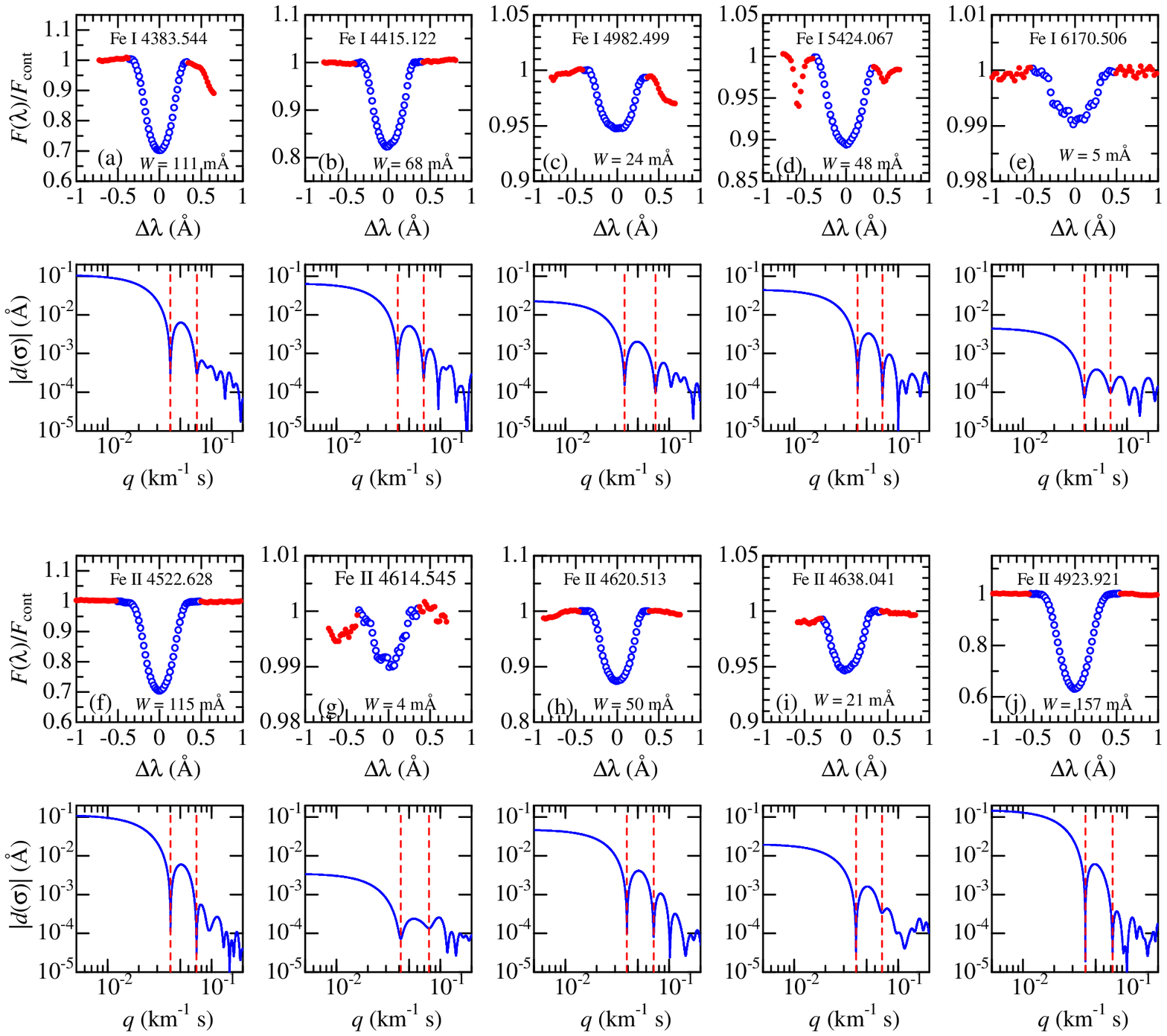}
\caption{
Each of the 10 sets (consisting of upper and lower panels) show 
the observed profiles and their Fourier transforms for the representative 
Fe~{\sc i} and Fe~{\sc ii} lines of different strengths.
Regarding the profile panel (where the normalised flux is plotted against the wavelength
shift relative to the line centre), the spectrum portion [$\lambda_{1}$, $\lambda_{2}$]
used to compute the Fourier transform is shown by blue open circles,
while the unused part is depicted by red dots. In the transform panel,
the Fourier transform amplitude ($|d(\sigma)|$) is displayed as function of the 
frequency (wavelength-independent $q$ is used here instead of $\sigma$)
and the positions of 1st- and 2nd-zeros are indicated by vertical dashed lines.
The upper sets are for 5 Fe~{\sc i} lines (Fe~{\sc i} 4383, 4415, 4982, 5424, 
and 6170), whereas the lower ones are for 5 Fe~{\sc ii} lines
(Fe~{\sc ii} 4522, 4614, 4620, 4638, and 4923). 
}
\label{fig4}
\end{minipage}
\end{figure*}

\subsection{Comparison with the classical case}

It is interesting to examine these $q_{1}$ and $q_{2}$ values derived for 571 lines
in context with the classical modelling using the rotational broadening function $G(\lambda)$
(see, e.g., Gray 2005) which is to be convolved with the intrinsic (unbroadened) thermal 
profile to simulate the observed (broadened) line profile. This formulation has a distinct 
merit especially in application of the Fourier technique, because the $v_{\rm e}\sin i$-dependent 
zero frequencies in $g(\sigma)$ (Fourier transform of the broadening function) are simply 
measured from the transform of the observed profile, because ``convolution'' in the real 
space turns into ``multiplication'' in the Fourier space. Actually, this way of profile 
modelling based on the broadening function is widely used in spectroscopic 
$v_{\rm e}\sin i$ determinations (like all 9 studies quoted in Table~1).

It should be stressed, however, that this simple convolution model is based on 
the postulation of line profile invariance over the stellar disk; i.e., 
$I(\lambda,\theta)/I_{\rm cont}(\theta)$ is assumed to be independent of $\theta$
($I$ is the specific intensity and $\theta$ is the angle between surface normal 
and the line of sight; e.g., $\theta =0$ at the disk centre), which does not hold 
in general and is thus nothing but an approximation.
This rotational broadening function $G(\lambda)$ is usually parametrised with 
$v_{\rm e}\sin i$ and $\epsilon$, where $\epsilon$ is the limb-darkening coefficient 
appearing in the limb-darkening relation of $I(\theta) = I(0)(1-\epsilon + \epsilon\cos\theta)$.  
The analytical expressions for the 1st- and 2nd-zero ($q_{1}$, $q_{2}$, and their ratio) 
in the Fourier transform of $G(\lambda)$ were already presented by Dravins et al. (1990) as
\begin{equation}
q_{1} = (0.610 + 0.062\epsilon + 0.027\epsilon^{2} + 0.012\epsilon^{3} + 0.004\epsilon^{4})/(v_{\rm e}\sin i),
\end{equation}
\begin{equation}
q_{2} = (1.117 + 0.048\epsilon + 0.029\epsilon^{2} + 0.024\epsilon^{3} + 0.012\epsilon^{4})/(v_{\rm e}\sin i),
\end{equation}
and
\begin{equation}
q_{2}/q_{1} = 1.831 - 0.108\epsilon - 0.022\epsilon^{2} + 0.009\epsilon^{3} + 0.009\epsilon^{4}.
\end{equation}
where $q_{1}$ and $q_{2}$ are in unit of km$^{-1}$s.
Regarding the limb-darkening coefficient $\epsilon$ in the present case of Sirius~A, 
the following relation was found to hold from the angle-dependence of the continuum\footnote{ 
What matters here is the $\epsilon_{\rm cont}$ derived from the 
$\theta$-dependence of $I_{\rm cont}$ at the continuum, because 
$\epsilon_{\lambda}$ (corresponding to $I_{\lambda}$ inside the line) 
turns out to be $\epsilon_{\rm cont}$ after all under the postulation 
of line profile invariance over the stellar disk mentioned above 
(which is required for the classical modeling based on the convolution 
of rotational broadeing function to be valid). That is, since this 
assumption requires that the residual intensity 
$I_{\lambda}(\theta)/I_{\rm cont}(\theta)$ be position-independent 
(i.e., free from $\theta$), the same $\theta$-dependence must hold 
for both $I_{\rm cont}$ and $I_{\lambda}$, which eventually yields 
$\epsilon_{\lambda} = \epsilon_{\rm cont}$.
} specific intensities calculated at various wavelengths: 
\begin{equation}
\epsilon = 1.0199 - 1.4956 \lambda_{\mu} + 0.6996 \lambda_{\mu}^{2},
\end{equation}
where $\lambda_{\mu}$ is in $\mu$m.
This relation is illustrated in Fig.~5, which suggests that $\epsilon$ progressive decreases 
with wavelength from $\sim 0.5$ (at $\sim$~4500~\AA) to $\sim 0.3$ (at $\sim$~7500~\AA).
Accordingly, for any given $v_{\rm e}\sin i$, the classical values of $q_{1}$ as well as $q_{2}$ (and 
also $q_{2}/q_{1}$) can be expressed as functions of $\lambda$ by combining Eq.~6 with Eq.~3--5.

\setcounter{figure}{4}
\begin{figure*}
\begin{minipage}{70mm}
\includegraphics[width=7.0cm]{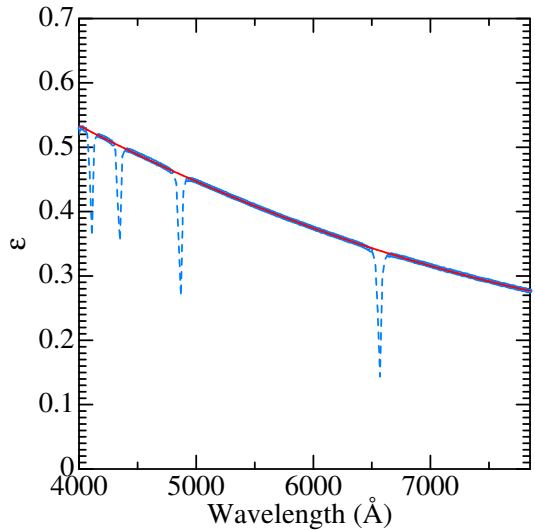}
\caption{
Wavelength-dependence of the linear limb-darkening coefficient $\epsilon$
(approximating the angle-dependence of specific intensity
as $I(\theta) = I(0) (1- \epsilon + \epsilon \cos\theta$) where $\theta =0$ 
for the disc centre), which was determined by applying the linear-regression
analysis to the theoretical $I(\theta)$ distribution calculated from the  
model atmosphere of Sirius~A ($T_{\rm eff} = 10000$~K, $\log g=4.30$, [M/H] = +0.5).
The result from the Balmer-line-included intensity is depicted by the blue dashed line
while that from the continuum intensity is shown by the blue symbols (which 
look as if a thick blue line because of being overlapped).
The 2nd-order polynomial determined by applying the least-squares fitting
to the latter continuum case (cf. Eq.~6) is also 
represented by the red solid line.
}
\label{fig5}
\end{minipage}
\end{figure*}

\setcounter{figure}{5}
\begin{figure*}
\begin{minipage}{80mm}
\includegraphics[width=8.0cm]{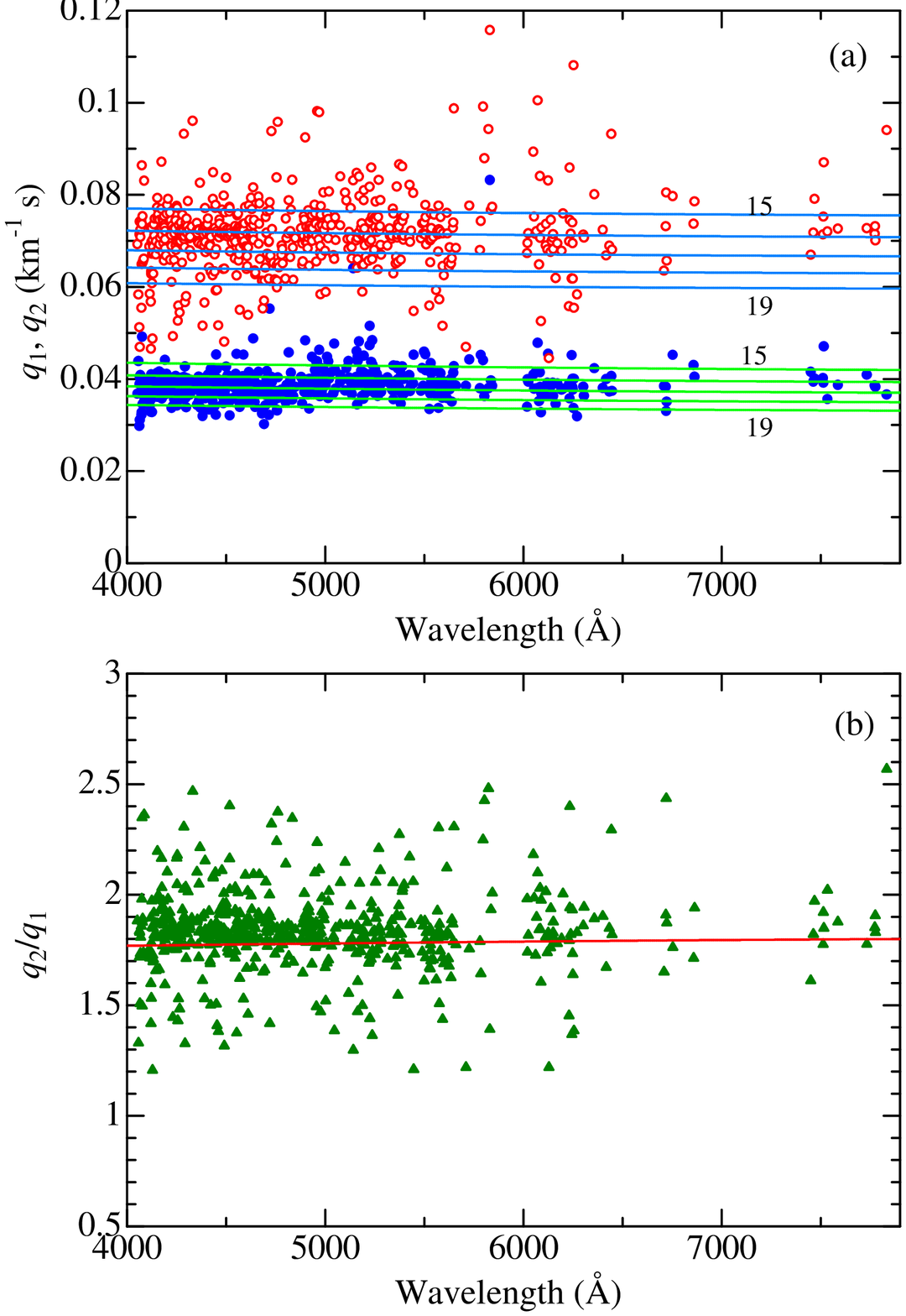}
\caption{
Observed 1st- and 2nd-zero frequencies of the Fourier transform amplitudes ($|d|$), 
which were calculated from the profiles of all 571 lines, plotted against
the wavelength: (a) $q_{1}$ and $q_{2}$, and (b) $q_{2}/q_{1}$. 
The classical relations $q_{1}(\lambda)$, $q_{2}(\lambda)$, and $q_{2}/q_{1}(\lambda)$
for $v_{\rm e}\sin i$ = 15, 16, 17, 18, and 19~km~s$^{-1}$, which were
derived from Eq.~3--6, are depicted by solid lines. 
}
\label{fig6}
\end{minipage}
\end{figure*}

In Fig.~6 are plotted the observed results of $q_{1}$, $q_{2}$, and $q_{2}/q_{1}$ against $\lambda$,
where the expected classical positions corresponding to 5 different $v_{e}\sin i$ values 
(15, 16, 17, 18, and 19~km~s$^{-1}$) are also indicated by lines.
Several characteristic trends can be read from Fig.~6.
\begin{itemize}
\item
As seen from the dispersion and number of outliers, $q_{2}$ (and thus $q_{2}/q_{1}$)
tends to suffer larger errors compared to $q_{1}$, which suggests that using $q_{1}$
is preferable while $q_{2}$ is not.   
\item
The classical $q_{1}$ or $q_{2}$ values (horizontal lines in Fig.~6a) are practically 
determined by $v_{\rm e}\sin i$ and do not appreciably depend upon wavelength, which 
means that the $\lambda$-dependence of $\epsilon$ (Fig.~5) is rather insignificant  
as long as the $\epsilon$ range of $\sim$~0.3--0.5 is concerned.   
\item
The average $v_{\rm e}\sin i$ values suggested from comparison of the observed zero frequencies 
with the classical predictions are not necessarily consistent between $q_{1}$ and $q_{2}$;
i.e., $\sim 17$~km~s$^{-1}$ (from $q_{1}$) and $\sim 16$~km~s$^{-1}$ (from $q_{2}$) (cf. Fig.~6a). 
Accordingly, the observed $q_{2}/q_{1}$ ratios (cf. Fig.~6b) tend to be slightly above $\sim 1.8$ 
(predicted by Eq.~5) on the average.
\end{itemize}

\section{Profile modelling of rotating star}

\subsection{Method and assumption} 

As seen from the results obtained in Sect.~2.4, there are inevitable limits and
uncertainties as long as the classical formulation is relied upon. In order to make 
a further step toward higher-precision $v_{e}\sin i$ determination or $v_{\rm e}$--$i$ 
separation, line profile modelling based on disk-integrated flux simulation of 
gravity-darkened rotating star is necessary, which adequately takes into account 
the position-dependent physical conditions by assigning many model atmospheres 
of different parameters.

Regarding the modelling of rotating star with gravity darkening, 
the same assumption and parameterisation as described in Sect. 2.1 and 2.2 of 
Takeda et al. (2008) were used: It is Roche model with rigid rotation, which 
is characterised by stellar mass ($M$), polar radius ($R_{\rm p}$), 
polar effective temperature ($T_{\rm eff,p}$), and equatorial rotational velocity 
($v_{\rm e}$), which suffice to express the surface quantities [$r(\Theta)$, 
$T_{\rm eff}(\Theta)$,\footnote{
The program CALSPEC assumes the relation 
$T_{\rm eff} \propto g^{\beta}$ in calculating $T_{\rm eff}(\Theta)$ 
from the effective gravity $g(\Theta$), where the exponent $\beta$ is 
evaluated according to the analytical formula $\beta = f(T_{\rm eff})$
constructed from Claret's (1998) Fig.~5 (cf. footnote~2 in Takeda et al. 
2008). In the present cases of sufficieltly high $T_{\rm eff}$ 
($\ga 9500$~K) where the stellar envelope is in radiative equilibrium, 
this formula yields $\beta = 0.25$ (von Zeipel's law) for all models.
Incidentally, Espinosa Lara \& Rieutord's (2011) more detailed 
theoretical simulation suggests that this classcal von Zeipel's value 
($\beta = 0.25$) is still a fairly good approximation unless 
rotation is very fast (cf. Fig.~4 therein).
}
 $g(\Theta)$, $v(\Theta)$] as functions of $\Theta$ 
(colatitude angle).
Since different model atmosphere corresponding to the local ($T_{\rm eff}$, $g$) 
is defined at each point on the surface, we can compute the flux profile
to be observed by an external observer by integrating the emergent  
intensity profile over the visible disk for any inclination angle ($i$) of 
the rotational axis, while using the program CALSPEC (see Sect. 4.1 in 
Takeda et al. 2008).
Accordingly, it is necessary to specify $M$, $R_{\rm p}$, $T_{\rm eff,p}$, 
$v_{\rm e}$, and $i$ as the fundamental model parameters.

\subsection{Adopted model parameters}

For the present case of Sirius~A, the reasonable parameters of $M = 2.1~{\rm M}_{\odot}$, 
$R_{\rm p} = 1.7~{\rm R}_{\odot}$, and $T_{\rm eff,p} = 10000$~K were adopted, 
which correspond to $\log g_{\rm p} = 4.30$. 

In order to examine the effect of changing $v_{\rm e}$, 10 models of different 
$v_{\rm e}$ (15, 30, 45, ... 135, 150~km~s$^{-1}$) were calculated,
while the $i$ values were so adjusted as to fix $v_{\rm e}\sin i$ at 15~km~s$^{-1}$.
The parameters of these models (numbered as 0, 1, 2, ... 9) are summarised
in Table~2, where the corresponding equatorial quantities ($\log g_{\rm e}$, $T_{\rm eff,e}$, 
and $R_{\rm e}$) are also presented.

\setcounter{table}{1}
\begin{table*}
\begin{minipage}{180mm}
\small
\caption{Models with different equatorial velocities.}
\begin{center}
\begin{tabular}{cccccccc}\hline
Model & $v_{\rm e}$ & $\sin i$ & $i$ &  $\log g_{\rm e}$ & $T_{\rm eff,e}$ & $R_{\rm e}$ & $v_{\rm e}/v_{\rm e}^{*}$ \\
number & (km~s$^{-1}$) &   & (deg) & (cm~s$^{-2}$) & (K) & (${\rm R}_{\odot}$)  & \\ 
\hline
 0 &  15 & 1.0000 & 90.0 &  4.298 & 9995 & 1.7008 & 0.031 \\
 1 &  30 & 0.5000 & 30.0 &  4.296 & 9981 & 1.7033 & 0.062 \\
 2 &  45 & 0.3333 & 19.5 &  4.291 & 9957 & 1.7073 & 0.093 \\
 3 &  60 & 0.2500 & 14.5 &  4.286 & 9923 & 1.7131 & 0.124 \\
 4 &  75 & 0.2000 & 11.5 &  4.278 & 9880 & 1.7205 & 0.155 \\
 5 &  90 & 0.1667 &  9.6 &  4.268 & 9826 & 1.7298 & 0.187 \\
 6 & 105 & 0.1429 &  8.2 &  4.257 & 9761 & 1.7408 & 0.219 \\
 7 & 120 & 0.1250 &  7.2 &  4.244 & 9687 & 1.7536 & 0.251 \\
 8 & 135 & 0.1111 &  6.4 &  4.228 & 9601 & 1.7684 & 0.284 \\
 9 & 150 & 0.1000 &  5.7 &  4.221 & 9504 & 1.7852 & 0.317 \\
\hline
\end{tabular}
\end{center}
These models were calculated for various combinations of ($v_{\rm e}$, $i$)
while $v_{\rm e}\sin i$ is fixed at 15~km~s$^{-1}$.
Other input parameters are common to all models:  $M =2.1$~M$_{\odot}$ (mass), 
$T_{\rm eff,p} = 10000$~K (polar effective temperature), and $R_{\rm p} = 1.7$~R$_{\odot}$ 
(polar radius). 
Shown in column 5--7 are the equatorial values of $\log g$, $T_{\rm eff}$, 
and $R$ (affected by rotation-induced centrifugal force). 
Column 8 gives $v_{\rm e}$-to-$v_{\rm e}^{*}$ ratio ($v_{\rm e}^{*} \equiv \sqrt{GM/R_{\rm e}}$
is the critical break-up velocity, where $G$ is the gravitational constant).
\end{minipage}
\end{table*}

As to the line profile calculation, the $A$ value (derived 
from $W$) was fixed as input abundance, and the atomic parameters taken from 
VALD were used. In addition, $v_{\rm t} = 2$~km~s$^{-1}$ (microturbulence) 
and [M/H] = +0.5 (metallicity of the local model atmosphere) were assumed as done 
in Sect.~2.2. 

\subsection{Simulated profiles and transforms}

In order to avoid complexity in interpreting the trend of theoretical results
and in their application to the observed data, the profile simulations were 
carried out only for 371 Fe lines (150 Fe~{\sc i} and 221 Fe~{\sc ii} lines) 
which make up the majority ($\sim 65\%$) of the 571 lines in total.
For each of the 371 lines, 10 kinds of flux profiles ($F(\lambda)$) for different $v_{\rm e}$ 
values (model~0, 1, ..., 9) were simulated by using CALSPEC. Then, the corresponding
Fourier transforms ($d(\sigma)$) were computed, from which the characteristic quantities
($q_{1}$, $q_{2}$, $h^{0}$, and $h^{1}$) were further measured, in the same way as
done in Sect.~2.3. The results are presented in ``fesimulated.dat'' of the online material.

As typical examples, Fig.~7 illustrates the computed profiles and their transforms 
(for $v_{\rm e}$ = 15, 60, 105, and 150~km~s$^{-1}$) as well as the $v_{\rm e}$-dependences 
of Fourier transform parameters ($q_{1}$, $q_{2}$, $h^{0}$, and $h^{1}$) for 4 representative 
lines (strong line and very weak line for both Fe~{\sc i} and Fe~{\sc ii}).
These figure panels suggest that variations in the line profile as well as the characteristic 
parameters of Fourier transform in response to changing $v_{\rm e}$ are appreciable only in 
the case of weak Fe~{\sc i} line, while quite marginal for the remaining cases of strong 
Fe~{\sc i} line and especially Fe~{\sc ii} lines (whichever weak or strong).

\setcounter{figure}{6}
\begin{figure*}
\begin{minipage}{160mm}
\includegraphics[width=16.0cm]{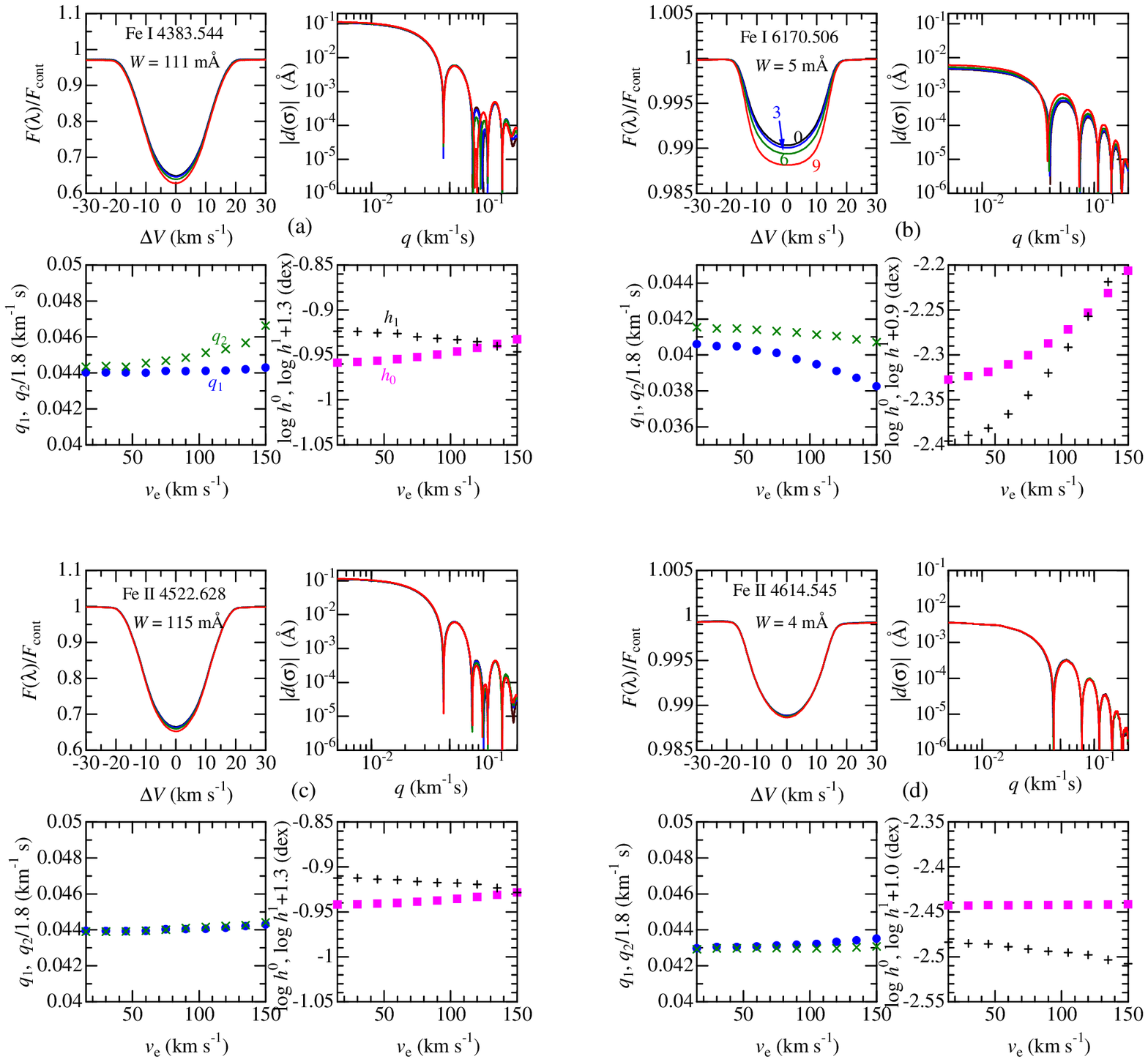}
\caption{Each figure set (consisting of 4 panels) illustrate the simulated line profiles
(for models 0, 3, 6, and 9) and their Fourier transform amplitudes, along with the 
$v_{\rm e}$-dependence of $q_{1}$ (1st zero; filled circles) and $q_{2}$ (2nd zero; 
St. Andrew's crosses) as well as of $h_{0}$ (main lobe height or equivalent width; 
filled squares) and $h_{1}$ (side lobe height; Greek crosses).
Note that the absolute scales of $q_{2}$ and $h_{1}$ are appropriately adjusted
for comparison purposes. (a) Fe~{\sc i} 4383 (strong Fe~{\sc i} line),
(b) Fe~{\sc i} 6170 line (very weak Fe~{\sc i} line, 
(c) Fe~{\sc ii} 4522 line (strong Fe~{\sc ii} line), and
(d) Fe~{\sc ii} 4614 line (very weak Fe~{\sc ii} line).
}
\label{fig7}
\end{minipage}
\end{figure*}

\section{Discussion}

\subsection{Temperature sensitivity of spectral lines}

As mentioned in Sect.~2.4, the conventional modelling of rotational broadening effect
(convolving the broadening function with the unbroadened intrinsic profile) 
is nothing but a simplified approximation. How the characteristics of line profiles 
are affected by stellar rotation depends on the physical properties of each line. 
Actually, the most important key parameter is the temperature sensitivity; that is, 
how the line strength reacts to a change of photospheric temperature.  
The reason why this parameter is significant is twofold:
(i) Since the line-forming depth shifts to upper lower-$T$ layer towards the limb,
contribution of the near-limb region to the line profile differs from line to line 
depending on the $T$-sensitivity.
(ii) As the rotation causes gravity darkening that lowers $T$ in the low-latitude region
near the equator (which corresponds to the near-limb region for the case of 
comparatively small $i$), this effect on the line profile is likewise 
$T$-sensitivity-dependent.

In a similar manner as adopted in Takeda \& UeNo (2017, 2019), the $T$-sensitivity indicator 
($K$) is defined as
\begin{equation} 
K \equiv {\rm d}\log W/{\rm d}\log T. 
\end{equation}
Practically, this parameter was numerically evaluated for each line as
\begin{equation}
K \equiv \frac{(W^{+100} - W^{-100})/W}{(+100-(-100))/10000},
\end{equation}
where $W^{+100}$ and $W^{-100}$ are the equivalent widths computed (with the 
same $A$ value derived from $W$) 
by two model atmospheres with only $T_{\rm eff}$ being perturbed by $+100$~K 
($T_{\rm eff} = 10100$~K) and  $-100$~K ($T_{\rm eff} = 9900$~K), respectively
(while other parameters are kept the same as the standard values; cf. Sect.~2.2).
The resulting $K$ vales for each of the 371 Fe lines are presented in 
``fesimulated.dat''.

An inspection of these results reveals a distinct difference in the range of $K$ between 
Fe~{\sc i} and Fe~{\sc ii} lines: $-15 \la K \la 0$ for the former while
$-5 \la K \la 5$ for the latter. This means that Fe~{\sc i} lines (negative $K$) tend 
to be weakened with an increase in $T$ (though its degree is considerably line-dependent), and 
that Fe~{\sc ii} lines ($K \sim 0$ on the average) are comparatively inert to a change in $T$.
The diversities of $K$ values observed in the same species are generally caused by the difference 
in the specific properties of each line such as $\chi_{\rm low}$ (lower excitation potential) 
or $W$ (line strength). It is important to note here, however, that the $K$ values of neutral 
Fe~{\sc i} lines are essentially determined only by $W$ (almost irrespective of $\chi_{\rm low}$). 
That is, weaker Fe~{\sc i} lines become progressively more $T$-sensitive with 
larger $|K|$ (cf. Fig.~A1d), indicating that very weak Fe~{\sc i} lines are most susceptible.
This explains why appreciable profile changes were observed in weak Fe~{\sc i} lines
(but not in  Fe~{\sc ii} lines or strong Fe~{\sc i} lines) in Fig.~7. 
In Appendix~A are separately discussed the behaviours of $K$ for Fe~{\sc i} and Fe~{\sc ii} 
lines in comparison with the solar case.

\subsection{Fourier transform parameters of simulated model profiles}

\setcounter{figure}{7}
\begin{figure*}
\begin{minipage}{130mm}
\begin{center}
\includegraphics[width=13.0cm]{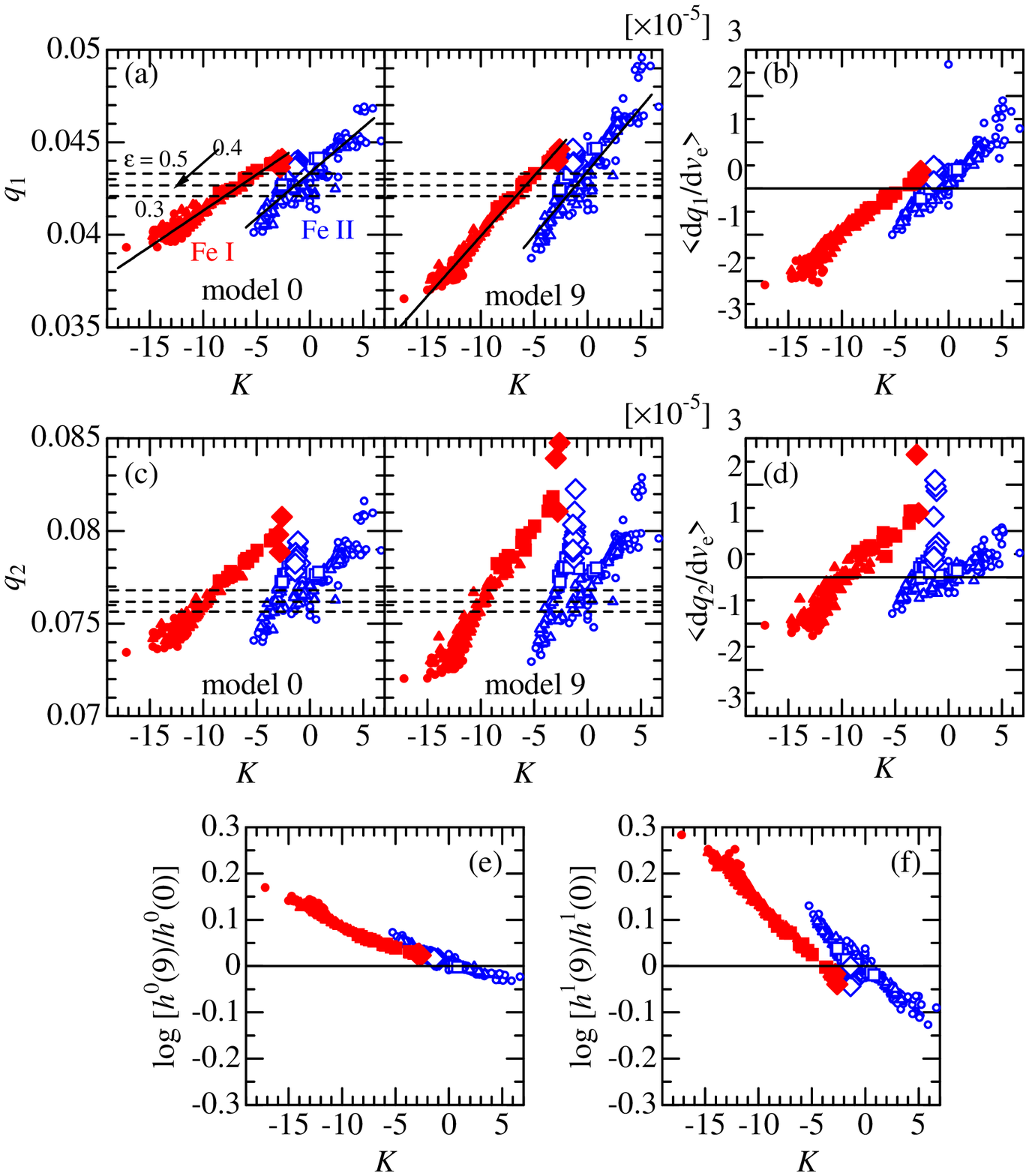}
\caption{
The top panels show the simulated behaviours of $q_{1}$ for the Fe~{\sc i} 
and Fe~{\sc ii} lines in terms of the relation to $K$ ($T$-sensitivity parameter) 
and its model-dependence ($v_{\rm e}$).  (a):  $q_{1}$ vs. $K$ plots for model~0 
($v_{\rm e}$ = 15~km~s$^{-1}$) and model~9 ($v_{\rm e}$ = 150~km~s$^{-1}$),
where filled and open symbols correspond to Fe~{\sc i} and Fe~{\sc ii} lines,
respectively (the size and shape of the symbols denote the difference in line
strengths; see the caption in Fig.~11 for more details). The linear regression lines derived 
from these $q_{1}$ vs. $K$ plots are also depicted by solid lines. The horizontal dashed lines 
represent the classical $q_{1}$ values (corresponding to $v_{\rm e}\sin i$ = 15~km~s$^{-1}$) 
for $\epsilon$ = 0.3, 0.4, and 0.5. 
(b) Mean gradients $\langle {\rm d}q_{1}/{\rm d}v_{\rm e}\rangle$ (km$^{-2}$s$^{2}$)
for Fe~{\sc i} and Fe~{\sc ii} lines, which were computed from the coefficients
of quadrature fit ($q_{1} = A + B v_{\rm e} + C v_{\rm e}^{2}$) as $B+2C \times (15+150)/2$
(i.e., ${\rm d}q_{1}/{\rm d}v_{\rm e}$ at the mid-$v_{\rm e}$), are plotted against $K$.
The middle panels (c and d) similarly display the simulated behaviours of $q_{2}$ for the Fe~{\sc i}
and Fe~{\sc ii} lines.
In the bottom panels (e) and (f) are plotted the model~9-to-0 ratios of $h^{0}$ 
(main lobe height) and $h^{1}$ (1st side lobe height) against $K$.
}
\label{fig8}
\end{center}
\end{minipage}
\end{figure*}

Fig.~8a and Fig~8c demonstrate how the Fourier zero frequencies ($q_{1}$ and $q_{2}$)
for the simulated profiles of Fe lines depend upon $K$ for two extreme $v_{\rm e}$ cases 
(model~0 ad model~9). Similarly, Fig.~8b and Fig.~8d show the mean derivatives 
with respect to $v_{\rm e}$ ($\langle {\rm d}q_{1}/{\rm d}v_{\rm e}\rangle$, 
$\langle {\rm d}q_{1}/{\rm d}v_{\rm e}\rangle$).
Further, the model~9-to-0 ratios of the main lobe height ($h^{0}$) and the first side lobe 
height ($h^{0}$) are also plotted in Fig.~8e and Fig.~8f.
The trends which can be read from these figures are summarised below, where the main focus
is placed on the behaviours of $q_{1}$ (to be used for comparison with the observational data).
\begin{itemize}
\item
Even for the case of negligible gravity darkening (model~0; left panel of Fig.~8a), 
the $q_{1}$ values can deviate from the classical prediction (0.042--0.043~km$^{-1}$s) by
$\delta q_{1}/q_{1} \sim \pm 0.1$ at most. Considering that 
$\delta v_{\rm e}\sin i/v_{\rm e}\sin i \sim \delta q_{1}/q_{1}$, intrinsic ambiguities of 
$\la$~1--2~km~s$^{-1}$ may be involved in the $v_{\rm e}\sin i$ determination of Sirius~A 
as long as the conventional method using rotational broadening function is invoked.  
\item
The $q_{1}$--$K$ plots for Fe~{\sc i} lines and Fe~{\sc ii} lines are clearly grouped
in two separate sequences, which show the same trend of ${\rm d}q_{1}/{\rm d}K > 0$ 
(i.e., $q_{1}$ progressively decreases with a decrease in $K$). 
Moreover, as the gravity darkening effect becomes appreciable with an increase in $v_{\rm e}$, 
the ${\rm d}q_{1}/{\rm d}K$ gradient gets steeper while the induced change of $q_{1}$ tending to be 
nearly proportional to $K$ ($\langle {\rm d}q_{1}/{\rm d}v_{\rm e}\rangle \propto K$; cf. Fig.~8b). 
\item
These behaviours can be reasonably explained in view of the difference in the contribution of 
the near-limb region to line profiles. Let us consider the typical case of significantly negative 
$K$ (i.e., weak Fe~{\sc i} line). Since the contribution from the region near the limb to  
the rotational broadening is relatively large in this case because the line strength increases 
there due to lowered $T$, the line profile becomes slightly wider which eventually moves
$q_{1}$ towards lower values. Accordingly, lines of negative $K$ tends to have comparatively 
lower $q_{1}$ for a given $v_{\rm e}\sin i$.
\item
The same argument holds also for the effect of increasing $v_{\rm e}$ that lowers $T$ 
in the low-latitude region near to the equator due to gravity darkening, 
which in effect leads to an increase of rotational broadening in the case of 
negative $K$ lines due to larger contribution of the cooler region near to the limb. 
Accordingly, a progressive increase of $v_{\rm e}$ leads 
to a decrease in $q_{1}$ (because the line profile becomes wider), an increase in $h^{1}$ 
(since the line tends to have a more rounded {\bf U}-shape) and an increase $h^{0}$ $(\equiv W)$ 
(because of the increased line strength near the limb) for such significantly $K < 0$ 
lines (i.e., weak Fe~{\sc i} lines), while $T$-insensitive lines of $K \sim 0$
are almost free from such effects. As such, the trends observed in Fig.~8 can be 
reasonably understood. 
\end{itemize} 

\subsection{Rotational velocity of Sirius~A}

Let us now address the main purpose of this study: getting information on $v_{\rm e}$ 
of Sirius~A by analysing the observed line profiles (Sect.~2.3) with the help of theoretical 
simulations for 371 Fe lines (Sect.~3.3). Only the Fourier 1st-zero frequency ($q_{1}$)
is employed for this analysis, since the 2nd zero ($q_{2}$) is less suitable because of
larger errors (cf. Sect.~2.4). 
The observed $q_{1}^{\rm obs}$ values of Fe lines are plotted against $K$ in Fig.~9.
This figure shows that these $q_{1}^{\rm obs}$ data for Fe~{\sc i} and Fe~{\sc ii} lines 
are distributed in separate two groups just like the theoretical prediction (cf. Fig.~8a),
though the dispersion is rather large.

\setcounter{figure}{8}
\begin{figure*}
\begin{minipage}{80mm}
\begin{center}
\includegraphics[width=8.0cm]{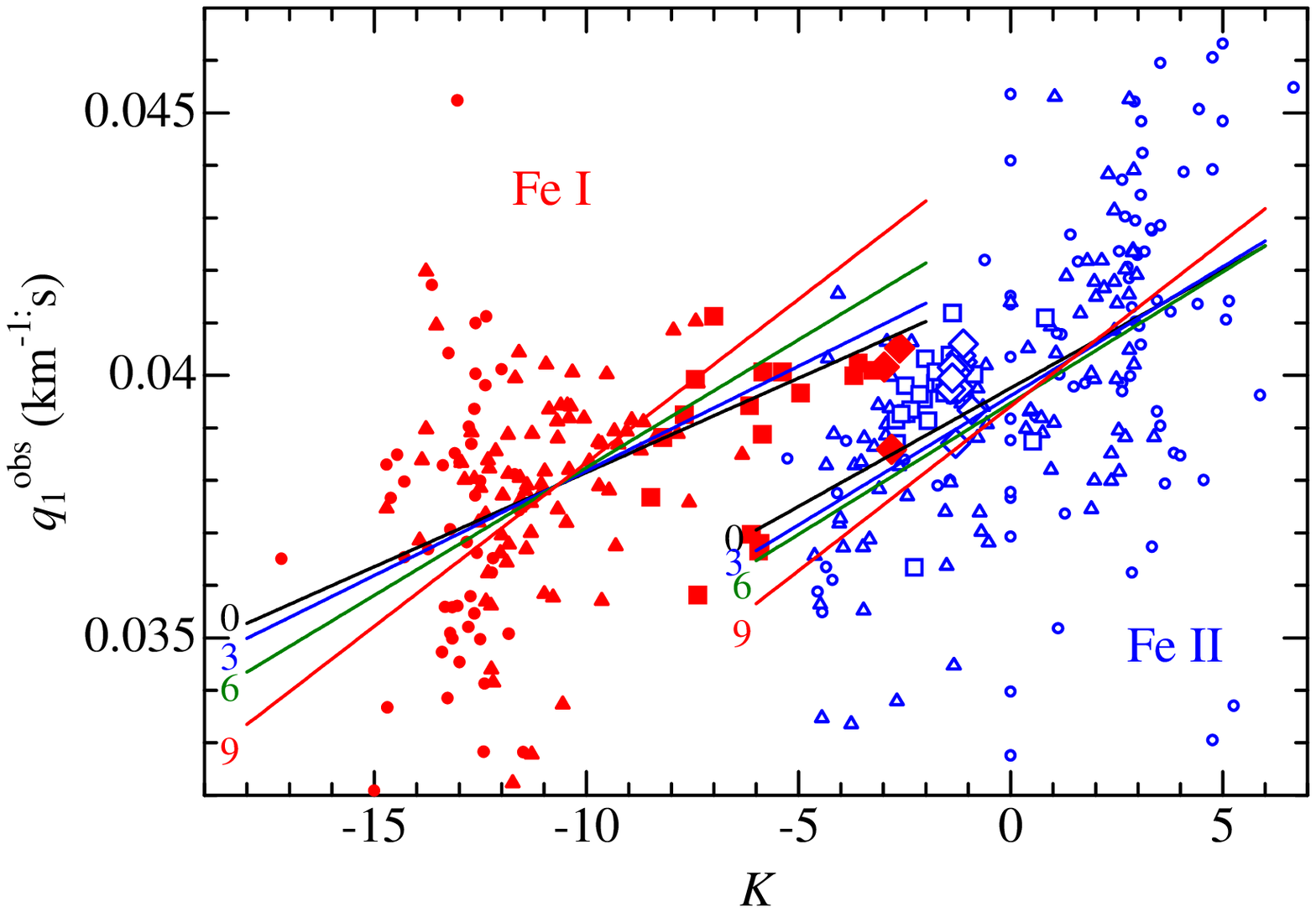}
\caption{
Observed $q_{1}$ values of Fe~{\sc i} and Fe~{\sc ii} lines plotted against $K$,
where the meanings of the symbols are the same as in Fig.~8.
The averaged trends (gradients) of theoretical $q_{1}$ vs. $K$ relations for 
models 0, 3, 6, and 9 (determined by linear-regression analysis; cf. Fig.~8a) 
are also depicted by the solid lines, which were multiplied by a scaling factor of 
$15/x^{*}$ in order to adjust the difference between the actual 
$v_{\rm e}\sin i (\equiv x)$ and the assumed value (15~km~s$^{-1}$) in the modelling
(see Table~3 for the $v_{\rm e}$-dependent values of $x^{*}$).
}
\label{fig9}
\end{center}
\end{minipage}
\end{figure*}

\setcounter{figure}{9}
\begin{figure*}
\begin{minipage}{120mm}
\begin{center}
\includegraphics[width=12.0cm]{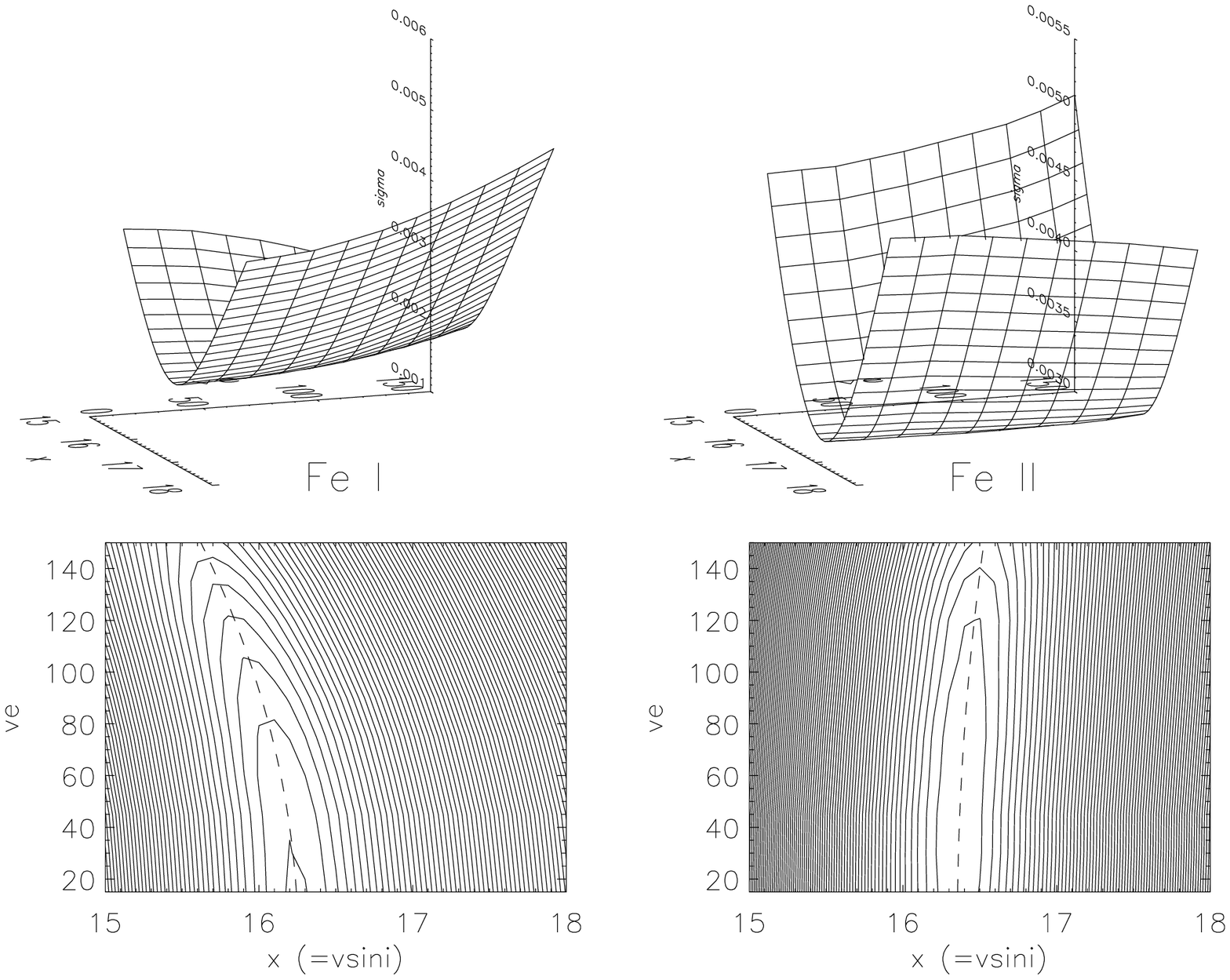}
\caption{
Graphical display of the behaviour of $\sigma$, which is the standard deviation 
between the simulated $q_{1}(x,v_{\rm e})$ (where $x\equiv v_{\rm e}\sin i$) and 
the observed $q_{1}^{\rm obs}$ for each of the Fe lines, where the results for 
Fe~{\sc i} and Fe~{\sc ii} lines are separately displayed in the left and 
right, respectively. Each set consists of the 3D representation of the $\sigma(x,v_{\rm e})$ 
surface (upper panel) and the contours of $\sigma$ on the $x$--$v_{\rm e}$ plane (lower panel).
The trace of trough bottom (connection of $x^{*}$ values at the minimum $\sigma$ for each given 
$v_{\rm e}$; cf. Table~3) is indicated by the dashed line in the contour plot.
}
\label{fig10}
\end{center}
\end{minipage}
\end{figure*}

In Sect.~2.3 were computed the theoretical $q_{1}^{{\rm th},j}$ values corresponding to 
10 models of different $v_{\rm e}^{j}$ ($j$ = 0, 1, ...,9) for each of the 371 Fe lines 
while fixing $v_{\rm e}\sin i$ at 15~km~s$^{-1}$ (cf. Table~2).
Note that, since the actual value of $v_{\rm e}\sin i$ (hereinafter denoted as $x$) is 
slightly different from 15~km~s$^{-1}$ (probably slightly larger; cf. Table~1),  
$q_{1}^{{\rm th},j}$ should be multiplied by a scaling factor of $15/x$ to adjust the difference. 
In effect, our task is to determine ($x$, $v_{\rm e}$) by comparing $q_{1}^{{\rm th},j}(15/x)$
with $q_{1}^{\rm obs}$ for a number of lines. For this purpose, the standard deviation 
$\sigma(x^{i},v_{\rm e}^{j})$ defined as
\begin{equation}
\sigma(x^{i},v_{\rm e}^{j}) \equiv 
\sqrt{ 
\frac{ \sum_{n=1}^{N} [q_{1}^{\rm obs}(n) - q_{1}^{{\rm th},j}(n)(15/x^{i})]^{2}}{N}.
}
\end{equation}
was computed for each combination of ($x^{i}$, $v_{\rm e}^{j}$),
where $x^{i} = 15.0 + 0.1i$ ($i$ = 0, 1. ..., 30) and $v_{\rm e}^{j} = 15 + 15j$ 
($j$ = 0, 1, ..., 9). Here, $n$ is the index of each line and $N$ is the total 
number of the lines used.
Since $\sigma$ is a measure of goodness-of-fit between observation and calculation, it may be 
possible to find the best ($x$, $v_{\rm e}$) solution by the requirement of minimising $\sigma$. 

Among the 371 Fe~{\sc i} lines, 13 lines showing considerable outlier $q_{1}^{\rm obs}$ values
($q_{1} < 0.032$ or $q_{1} > 0.047$) were excluded. In addition, 6 very weak lines of $W\sim$~1--3~m\AA\
(Fe~{\sc i} 4265, 4515, 4800, 6752; Fe~{\sc ii} 4559, 4975) could not be used because their 
theoretical $q_{1}^{{\rm th},j}$ turned out to show some numerical instabilities.
Accordingly, eliminating 19 lines (these unused lines are marked with `X' in ``fesimulated.dat''), 
the remaining 352 lines were finally used for the analysis, where 145 Fe~{\sc i} lines and 
207 Fe~{\sc ii} lines were separately treated.
The resulting behaviours of $\sigma$ (3D surface and contour plots) are displayed in Fig.~10 
(where the results for Fe~{\sc i} and Fe~{\sc ii} are shown in the left and right panels, 
respectively), and the characteristics of the trough in the $\sigma$ surface are summarised 
in Table~3, from which the following results can be concluded.
\begin{itemize}
\item
Although the best solution of ($x$, $v_{\rm e}$) may be found by searching for the local 
minimum of $\sigma$, consistent results could not be obtained in this way: For the case of 
Fe~{\sc i}, $\sigma_{\rm I}$ monotonically decreases with a decrease in $v_{\rm e}$ 
along the trough until the lower boundary of $v_{\rm e}$ (15~km~s$^{-1}$) is reached. 
On the other hand, $\sigma_{\rm II}$ is insensitive to a change of $v_{\rm e}$
along the trough. Admitting that a very shallow minimum of $\sigma_{\rm II}$ is superficially 
seen at $v_{\rm e} \sim 30$~km~s$^{-1}$ (cf. Table~3), this can not be regarded as a robust solution.
\item
It would be more useful to pay attention to the trace line connecting ($x^{*}$, $v_{\rm e}$) 
where $x^{*}$ corresponds to the minimum of $\sigma$ trough for each given $v_{\rm e}$. 
This tracing (cf. Table~3 for the data) is shown by the dashed line in the contour plot (Fig.~10).
Though these trace lines for Fe~{\sc i} and Fe~{\sc ii} do not intersect with each other, 
the closest approach is again at the boundary of $v_{\rm e} = 15$~km~s$^{-1}$, where
the difference between $x^{*}_{\rm I}$ and $x^{*}_{\rm II}$ is 0.012~km~s$^{-1}$
(cf. Table~3).
\item
Here, it is worthwhile to estimate errors involved in $x^{*}$.
Intuitively speaking, the scaling factor $15/x^{*}$ in Eq.~9 (by which 
$q_{1}^{\rm th}$ is to be multiplied) is so determined as to adjust the 
location (i.e., height) of the linear regression line (representing 
the ${\rm d}q_{1}^{\rm th}/{\rm d}K$ gradient), so that it passes through 
the dispersion of $q_{1}^{\rm obs}$ as impartially as possible 
(cf. Fig.~9).
Regarding the distribution of $q_{1}^{\rm obs}$, the precision in 
determining the dispersion centre may be regarded as $\sigma/\sqrt{N}$ 
(mean error denoted as $\delta q_{1}$) by using $\sigma$ defined in Eq.~9,
which makes $\delta q_{1} \sim 0.0002$~km$^{-1}$s (for both Fe~{\sc i}
and Fe~{\sc ii}) according to the $\sigma$ values presented in Table~3.
If the mean error in $x^{*}$ (corresponding to $\delta q_{1}$) is denoted 
as $\delta x^{*}$, the relation 
$\delta x^{*}/x^{*} \sim \delta q_{1}/q_{1}$ holds
(because $q_{1} \propto 1/x^{*}$). 
Putting $\delta q_{1} \sim 0.0002$~km$^{-1}$s, $q_{1} \sim 0.04$~km$^{-1}$s 
(cf. Fig.~9), and $x^{*} \sim 16$~km~s$^{-1}$ (cf. Table~3) into this 
relation, we obtain $\delta x^{*} \sim 0.08$~km~s$^{-1}$ for both Fe~{\sc i} 
and Fe~{\sc ii}.
\item
Since the $x^{*}$ values suffer uncertainties of $\sim 0.08$~km~s$^{-1}$ for both Fe~{\sc i} 
and Fe~{\sc ii}, $x^{*}_{\rm I}$ and $x^{*}_{\rm II}$ may be regarded as practically 
equivalent as long as the difference is $\la 0.16$~km~s$^{-1}$ (i.e., within expected errors).
So, an inspection of Table~2 suggests that the acceptable upper limit of $v_{\rm e}$ 
is around $\sim$~30--40~km~s$^{-1}$ (above which $x^{*}_{\rm I}$ and $x^{*}_{\rm II}$ 
are inconsistent and thus should be ruled out).
This, in turn, establishes the final $v_{\rm e}\sin i$ solution (consistent $x^{*}$ value
for both Fe~{\sc i} and Fe~{\sc ii}) as 16.3~km~s$^{-1}$ 
with a precision of $\sim 0.1$~km~s$^{-1}$ (see the discussion above on the errors).
Further, by combining the results of $16 \le v_{\rm e} \la$~30--40~km~s$^{-1}$ and 
$v_{\rm e}\sin i = 16.3$~km~s$^{-1}$, information on the possible range of $i$ may be 
obtained as $25^{\circ} \la i \le 90^{\circ}$.
\end{itemize}

\setcounter{table}{2}
\begin{table*}
\begin{minipage}{180mm}
\small
\caption{Behaviours of $\sigma$ trough for Fe~{\sc i} and Fe~{\sc ii} lines.}
\begin{center}
\begin{tabular}{ccccc}\hline
$v_{\rm e}$ & $x^{*}_{\rm I}$ & $x^{*}_{\rm II}$ & $\sigma^{*}_{\rm I}$ &  $\sigma^{*}_{\rm II}$ \\
(km~s$^{-1}$) & (km~s$^{-1}$) & (km~s$^{-1}$) & (km$^{-1}$s) & (km$^{-1}$s) \\ 
\hline
   15 &  16.2399 & 16.3562 & 0.0018324 & 0.0030313 \\
   30 &  16.2187 & 16.3620 & 0.0018396 & 0.0030270 \\
   45 &  16.1939 & 16.3651 & 0.0018444 & 0.0030273 \\
   60 &  16.1430 & 16.3821 & 0.0018551 & 0.0030283 \\
   75 &  16.0887 & 16.3995 & 0.0018756 & 0.0030282 \\
   90 &  16.0164 & 16.4201 & 0.0019006 & 0.0030339 \\
  105 &  15.9306 & 16.4393 & 0.0019374 & 0.0030412 \\
  120 &  15.8345 & 16.4606 & 0.0019813 & 0.0030473 \\
  135 &  15.7360 & 16.4923 & 0.0020401 & 0.0030643 \\
  150 &  15.6284 & 16.5289 & 0.0021104 & 0.0030912 \\
\hline
\end{tabular}
\end{center}
These data show the characteristics of the trough in the $\sigma(x,v_{\rm e})$ surface 
($x \equiv v_{\rm e}\sin i$) defined by Eq.~9 for each group of Fe~{\sc i} and Fe~{\sc ii} lines.
$x^{*}$ is the $x$ value at the minimum $\sigma(x,v_{\rm e})$ for each given $v_{\rm e}$,
and $\sigma^{*}$ is the corresponding $\sigma(x^{*},v_{\rm e})$. The trace of $x^{*}$ as a
function of $v_{\rm e}$ is shown by the dashed line in the contour plot of Fig.~10.
\end{minipage}
\end{table*}

\subsection{Implication of the results}

Finally, some discussion on the rotational feature of Sirius~A obtained in Sect.~4.3 may 
be due in connection with the published studies and from the viewpoint of 
the Sirius system as a whole.

The $v_{\rm e}\sin i$ value of $16.3 (\pm 0.1)$~km~s$^{-1}$ is favourably compared with previous 
determinations (mostly in the range of 15~$\la v_{\rm e}\sin i \la 17$~km~s$^{-1}$;
especially recent ones are around $\simeq 16.5$~km~s$^{-1}$), which indicates that the 
conventional method using the classical rotational broadening function is still useful and 
sufficient unless very high precision is required. 

In this study, precise separation of $v_{\rm e}$ and $i$ was not feasible, and
only the possible ranges could be subtended as $16 \le v_{\rm e} \la$~30--40~km~s$^{-1}$ 
and $25^{\circ} \la i \le 90^{\circ}$. It is worth discussing these ranges in connection 
with the direction angle ($i_{\rm orb}$) of the orbital motion, which is accurately
established as $i_{\rm orb}= 136.336^{\circ} \pm 0.040^{\circ}$ (Bond et al. 2017)
or $i_{\rm obs} = 43^{\circ}$ $( = 180^{\circ} - 136^{\circ})$ if the range is restricted 
to $0^{\circ}$--$90^{\circ}$ by neglecting the sign of orientation.
If the rotational axis is strictly aligned with the axis of orbital motion 
($i = i_{\rm orb} = 43^{\circ}$), $v_{\rm e}$ would have a value of 
$16.3/\sin 43^{\circ}$ = 24~km~s$^{-1}$, which is almost in-between the 
$v_{\rm e}$ range concluded in this study. Accordingly, it is likely that the alignment is
nearly accomplished between the rotational axis of Sirius~A and the orbit axis of Sirius system. 
It would be worthwhile to examine whether a magnetic field variation exists with a modulation 
period of $\sim$~3--4~d (rotation period of 1.7~R$_{\odot}$ star with $v_{\rm e} \sim 24$~km~s$^{-1}$ 
is $\sim 3.6$~d) by high-precision spectropolarimetric observations such as done by Petit et al. (2011).   

The consequence of $16 \le v_{\rm e} \la$~30--40~km~s$^{-1}$ corroborates that Sirius~A is
an intrinsically slow rotator, which indicates that this star is atypical among early-A dwarfs 
(many of them rotate with $v_{\rm e} \sim$~100--300~km~s$^{-1}$) and belongs to the minority 
group of slowly-rotating stars. This picture is quite consistent with the results of previous 
spectroscopic studies that this star shows distinct chemical anomalies (Am phenomena) 
which are considered to be triggered by slow rotation.
  
From the theoretical point of view, it is not straightforward to understand why 
Sirius~A rotates so slowly. Since a considerable fraction of Am stars are known to be 
spectroscopic binaries (of comparatively short period), deceleration due to tidal 
interaction is believed to be an important mechanism for slow rotation 
which eventually gives rise to chemical anomaly. However, since Sirius is a wide binary 
system of 50~yr period, it is not clear whether an efficient tidal braking was operative, 
even if both components could have been much closer in the past (e.g., the period as well as 
the size of orbit would have been by $\sim 5$ times smaller when the system was born; cf. 
Bond et al. 2017). 

As a matter of fact, since the large orbital eccentricity ($e \sim 0.6$) is rather 
unusual, it is still in dispute whether and how both components have physically interacted 
with each other to result in the current system, and different theoretical models 
have been proposed; e.g., evolution of $e$ due to impulse by episodic mass loss at the 
time of periastron (Bona\u{c}i\'{c} Marinovi\'{c}, Glebbeek \& Pols 2008) or star ejection 
from the triple system due to orbital instability to leave a binary system of large $e$
(Perets \& Kratter 2012). The consequence of this study on the rotational properties of Sirius~A
(intrinsically slow rotation, possible alignment between the axes of rotation and orbital motion)
may serve as observational constraints for such theoreticians intending to develop evolutionary
models of Sirius.  

\section{Summary and conclusion}

Sirius~A (classified as A1V) is known to be rather unusual for its apparent sharp-line 
nature ($v_{\rm e}\sin i \la $~20~km~s$^{-1}$), in contrast to many A-type main-sequence 
stars showing large $v_{\rm e}\sin i$ values (typically $\sim$~100--300~km~s$^{-1}$).

This may remind of the case of Vega (A0V) similarly showing low $v_{\rm e}\sin i$
($\sim 20$~km~s$^{-1}$), which is nothing but a superficial effect of very small
$i$ and its $v_{\rm e}$ is actually large ($\sim 200$~km~s$^{-1}$). 

Unlike Vega, however, Sirius~A is suspected to be an intrinsic slow rotator,
because it shows distinct surface chemical peculiarities (Am star) which 
are typically seen in slowly rotating stars of low $v_{\rm e}\sin i$..
For all that this is a reasonable surmise, it is important to confirm it
observationally, which was the motivation of this study. 

Towards the intended goal of getting information on $v_{\rm e}$ of Sirius~A 
separated from the inclination effect, a detailed profile analysis 
using the Fourier transform technique was conducted for hundreds of 
spectral lines, where the line-by-line differences in the centre--limb 
behaviours and the gravity darkening effect were explicitly taken into 
account based on the simulated rotating-star models.

The simulations showed that the $v_{\rm e}$-dependence of $q_{1}$ (1st zero
frequency of Fourier transform useful for detecting delicate profile difference) 
is essentially determined by the $T$-sensitivity parameter ($K$) 
differing from line to line, and that Fe~{\sc i} lines (especially those 
of very weak ones) are more sensitive to $v_{\rm e}$ than Fe~{\sc ii} lines.   
By comparing the theoretical and observed $q_{1}$ values for a number of 
Fe~{\sc i} and Fe~{\sc ii} lines, the following results were obtained.

The $v_{\rm e}\sin i$ value for Sirius~A could be established with sufficient precision 
as $16.3 (\pm 0.1)$~km~s$^{-1}$  by requiring that both Fe~{\sc i} and Fe~{\sc ii} 
lines yield consistent results. This is reasonably consistent with previous 
determinations (mostly in the range of 15~$\la v_{\rm e}\sin i \la 17$~km~s$^{-1}$;
especially recent ones are around $\simeq 16.5$~km~s$^{-1}$).

Although precise separation of $v_{\rm e}$ and $i$ was not feasible, $v_{\rm e}$ was
concluded to be within  $16 \le v_{\rm e} \la$~30--40~km~s$^{-1}$,
which further results in  $25^{\circ} \la i \le 90^{\circ}$.
Since the ($v_{\rm e}$, $i$) values of (24~km~s$^{-1}$, 43$^{\circ}$), which are
expected if the rotational and orbital axes are aligned, are almost in-between
these ranges, it is possible that such an alignment is realised. 

Now that the nature of intrinsically slow rotator has been established for Sirius~A,
its surface abundance anomaly is naturally understood in context of the general 
characteristics of upper main-sequence stars (i.e., advent of chemical peculiarities
is accompanied by slow rotation). The consequences of this study may serve as 
observational constraints for modelling the evolution of Sirius system.

\section*{Acknowledgments}
This research has made use of the SIMBAD database, operated by CDS, 
Strasbourg, France. 
This work has also made use of the VALD database, operated at Uppsala 
University, the Institute of Astronomy RAS in Moscow, and the University of Vienna.

\section*{Data availability}

The data underlying this article are available in the 
online supplementary material.

\appendix
\section{Line-dependence of $T$-sensitivity parameter}

As clarified in Sect.~4.2, the behaviours of Fourier transform parameters (such as 
zero frequencies) characterising the spectral line profiles are essentially controlled 
by the temperature sensitivity parameter ($K$) defined by Eq.~7. Then, how this $K$ 
depends upon the properties of individual lines?  
Here, the case of the Sun (early G dwarf) is explained at first before going to  
Sirius~A, because contrasting these two is useful for understanding the trends.

\subsection{Solar case}

According to Takeda \& UeNo (2017, 2019), three factors are involved in determining 
the $K$ values of Fe lines for the solar case: (i) ionisation degree (neutral or ionised),
(ii) lower excitation potential ($\chi_{\rm low}$), and (iii) line strength (equivalent width $W$).
Above all important is (i): Since most Fe atoms are in the once-ionised stage (Fe~{\sc ii})
and only a small fraction remain neutral (Fe~{\sc i}) in the line-forming region
of the solar atmosphere (cf. Fig.~A1e), Fe~{\sc i} and Fe~{\sc ii} may be called minor- and
major-population species, respectively. 
Under this situation, the $T$-dependence for the number population ($n_{i}$) of level $i$ 
can be expressed with the help of the Boltzmann--Saha equation as
\begin{equation}
n_{i}^{\rm I} \propto n_{0}^{\rm II} \exp{[(\chi_{\rm ION}^{\rm I}-\chi_{i}^{\rm I})/(kT)]} 
\end{equation}
and
\begin{equation}
n_{i}^{\rm II} \propto n_{0}^{\rm II} \exp{[-\chi_{i}^{\rm II}/(kT)]}
\end{equation}
where $n_{0}^{\rm II}$ is the ground level population of Fe~{\sc II} insensitive to $T$
(because of the major population species), $\chi_{i}$ is the excitation potential of level $i$,
$\chi_{\rm ION}$ is the ionisation potential (from the ground level), and $k$ is the Boltzmann
constant (superscript `I' and `II' correspond to Fe~{\sc i} and Fe~{\sc ii}, respectively). 

Accordingly, in the weak-line case where $W$ is almost proportional to the number 
population ($n$) of the lower level, the $T$-dependence of $K$ is approximately expressed as
\begin{equation}
K^{\rm I} \sim -11604 \, (\chi_{\rm ION}^{\rm I} -\chi_{\rm low}^{\rm I})/T \; (<0)
\end{equation} 
and 
\begin{equation}
K^{\rm II} \sim +11604 \, \chi_{\rm low}^{\rm II}/T \; (>0)
\end{equation}
where $\chi_{\rm low}$ and $\chi_{\rm ION}$ are in unit of eV and $T$ is in K 
(cf. Sect~4.1 in Takeda \& UeNo 2017).
As a matter of fact, it can be seen from Fig.~A1a that the trend of $K$ for
weak lines (blue small symbols) roughly follow these two relations.
Then, as lines get stronger and more saturated, $W$ is not proportional to $n$
any more and its $T$-sensitivity (or relative variation of $W$ for a change in $T$) 
becomes weaker than that given by Eq.~A3 and A4, which explains why stronger lines 
of larger $W$ tend to show smaller $|K|$ values compared to weak lines (cf. Fig.~A1c).
However, the $K$ vs. $W$ trend in Fig~A1c is not necessarily tight because $K$ depends 
(not only $W$ but) also on $\chi_{\rm low}$.

\subsection{Case of Sirius~A}

Although the $K$ vs. $\chi_{\rm low}$ plot in Fig.~A1b appears qualitatively similar 
to that of the solar case (Fig.~A1a), the situation is rather different.
Since (a) the sign of $K$ for Fe~{\sc ii} can be negative as well as positive 
and (b) the $K$ vs. $\chi_{\rm low}$ trend for each group of similar line strengths 
is not so tight (i.e., larger dispersion as compared to Fig.~A1a), it is apparent that 
Eq.~A3 and A4 are not directly applicable to A-type stars.

The reason is that the fundamental postulation of Fe~{\sc ii} being the major population 
species (i.e., dominant fraction in the line-forming region) does not hold any more 
as illustrated in Fig.~A1f, which shows that fractions of Fe~{\sc ii} and Fe~{\sc iii} 
are comparable around $\tau_{5000} \sim 1$. Accordingly, unlike the case for the Sun, 
describing the trend of $K$ by simple analytical expressions (in application of the 
Saha--Boltzmann equation) is rather difficult.

Nevertheless, a simple and useful relation for Fe~{\sc i} lines can be read 
from Fig.~A1d, which shows that $K^{\rm I}$ vs. $\log W$ plots (filled symbols) are 
fairly tight. This means that $K^{\rm I}$ is practically only $W$-dependent 
(regardless of $\chi_{\rm low}$), which may be interpreted as follows.

Let us consider the atmospheric layer where Fe~{\sc iii} is almost the dominant ionisation 
stage, while Fe~{\sc i} as well as Fe~{\sc ii} are minor population species.
Although Eq.~A1 still holds in this case, $n_{0}^{\rm II}$ is no more insensitive to $T$ 
but can be written (similarly to Eq.~A1) by using $n_{0}^{\rm III}$ as
\begin{equation}
n_{0}^{\rm II} \propto n_{0}^{\rm III} \exp{[(\chi_{\rm ION}^{\rm II}-0)/(kT)]}. 
\end{equation}
Inserting Eq.~A5 into Eq.~A1, we obtain
\begin{equation}
n_{i}^{\rm I} \propto n_{0}^{\rm III} \exp{[(\chi_{\rm ION}^{\rm I} + \chi_{\rm ION}^{\rm II} - \chi_{i}^{\rm I})/(kT)]},
\end{equation}
where $n_{0}^{\rm III}$ is inert to changing $T$ because Fe~{\sc iii} is the major population.
Since $\chi_{i}^{\rm I}$ (ranging 0--5~eV in the present case; cf. Fig.~A1b) is
relatively insignificant compared to $\chi_{\rm ION}^{\rm I}$ + $\chi_{\rm ION}^{\rm II}$ 
of $\sim 24$~eV (= 7.87 + 16.18~eV), Eq.~A6 indicates that $n_{i}^{\rm I}$ is insensitive to 
a change in $\chi_{i}^{\rm I}$, which may explain why $K^{\rm I}$ essentially
depends only on $W$.

\setcounter{figure}{0}
\begin{figure*}
\begin{minipage}{100mm}
\begin{center}
\includegraphics[width=10cm]{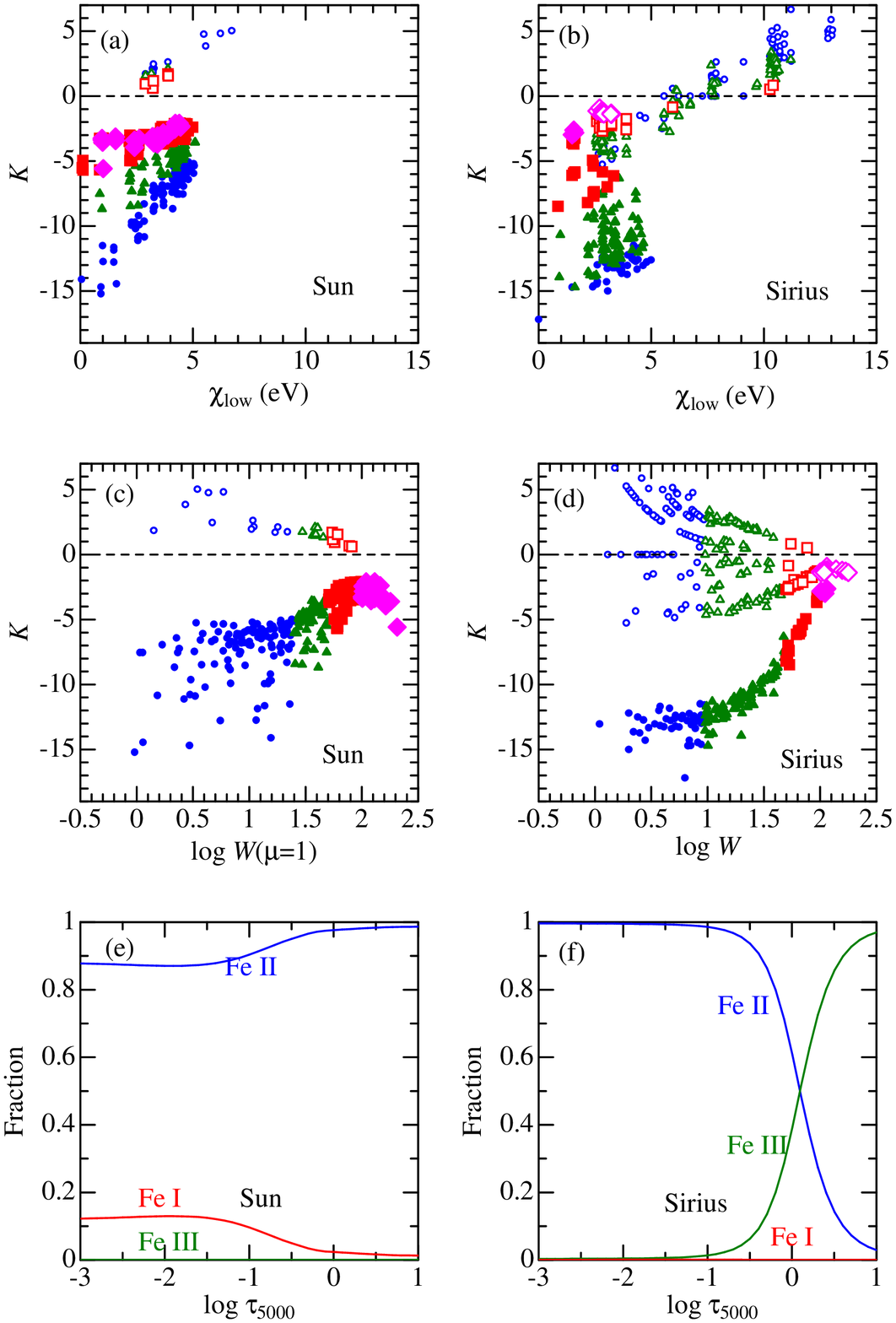}
\caption{
Panels (a)--(d) show how the $K$ values ($T$-sensitivity indicator) evaluated
for various  Fe lines depends upon $\chi_{\rm low}$ (lower excitation potential; 
top panels) and $W$ (equivalent width in m\AA; middle panels) for the case of Sun 
(left panels; 280 Fe~{\sc i} and 26 Fe~{\sc ii} lines; cf. Takeda \& UeNo 2019) and 
Sirius (right panels; 150 Fe~{\sc i} and 221 Fe~{\sc ii} lines adopted in this study).
The filled and open symbols correspond to Fe~{\sc i} and Fe~{\sc ii} lines, 
respectively. Lines of different strengths classes are discriminated by 
the shape and the size (larger for stronger lines) of symbols:
circles (blue) $\cdots$ $W <$~25~m\AA, 
triangles (green) $\cdots$ 25~m\AA~$\le W <$~50~m\AA,
squares (pink) $\cdots$ 50~m\AA~$\le W <$~100~m\AA, and
diamonds (brown) $\cdots$ 100~m\AA~$\le W$.
In the bottom panels (e) and (f) are shown the the number population fractions 
of Fe atoms in neutral (Fe~{\sc i}), once-ionised (Fe~{\sc ii}), and twice-ionised 
(Fe~{\sc iii}) stages as functions of the standard continuum optical depth at 5000~\AA,  
which were computed from Kurucz's (1993) model atmospheres for the Sun (e) and Sirius (f). 
}
\label{fig11}
\end{center}
\end{minipage}
\end{figure*}

\end{document}